\documentclass[a4paper,fleqn,usenatbib]{mnras}

\usepackage{newtxtext,newtxmath}
\usepackage[T1]{fontenc}
\usepackage{ae,aecompl}

\usepackage{graphicx}	% Including figure files
\usepackage{amsmath}	% Advanced maths commands
\usepackage{amssymb}	% Extra maths symbols

\title[X-ray Spectra and Light Curves]{X-ray Spectra and Light Curves of Cooling Novae and a Nova-Like}

\author[B. Sun et al.]{Bangzheng Sun,$^{1}$\thanks{E-mail: bsun49@wisc.edu}
Marina Orio,$^{1,2}$\thanks{E-mail: orio@astro.wisc.edu}
Andrej Dobrotka,$^{3}$
Gerardo Juan Manuel Luna,$^{4,5,6}$
\newauthor
Sergey Shugarov$^{7,8}$
and Polina Zemko$^{9}$
\\
$^{1}$Department of Astronomy, University of Wisconsin-Madison, Madison, WI 53715, U.S.A.\\
$^{2}$INAF - Padova Observatory, vicolo Osservatorio, 5 - 35122 Padova, Italy\\
$^{3}$Advanced Technologies Research Institute, Faculty of Materials Science and Technology in Trnava,
 Slovak University of Technology in Bratislava,\\
 Bottova 25, 917 24, Trnava, Slovakia \\
$^{4}$CONICET - ACONICET-Universidad de Buenos Aires, Instituto de Astronom\'ia y F\'isica del Espacio (IAFE), Av. Inte. G\"uiraldes 2620,
C1428ZAA,\\ Buenos Aires, Argentina\\
$^{5}$Universidad de Buenos Aires, Facultad de Ciencias Exactas y Naturales, Buenos Aires, Argentina \\
$^{6}$ Universidad Nacional de Hurlingham, Av. Gdor. Vergara 2222, Villa Tesei, Buenos Aires, Argentina \\
$^{7}$Astronomical Institute of the Slovak Academy of Sciences,
059 60 Tatransk'a Lomnica, The Slovak Republic
\\
$^{8}$ P.K. Sternberg Astronomical Institute, M.V. Lomonosov Moscow State
University, Russia\\
$^{9}$University of Padova, 35122 Padova, Italy } 

\date{Accepted XXX. Received YYY; in original form ZZZ}

\pubyear{2020}

\begin{document}
\label{firstpage}
\pagerange{\pageref{firstpage}--\pageref{lastpage}}
\maketitle

\begin{abstract}
We present X-ray observations of novae V2491 Cyg and KT Eri about 9 years post-outburst, of the dwarf nova and post-nova candidate EY Cyg, and of a VY Scl variable. The first three objects were observed with XMM-Newton, KT Eri also with the Chandra ACIS-S camera, V794 Aql with the Chandra ACIS-S camera and High Energy Transmission Gratings. The two recent novae, similar in outburst amplitude and light curve, appear very different at quiescence. Assuming half of the gravitational energy is irradiated in X-rays, V2491 Cyg is accreting at $\dot{m}=1.4\times10^{-9}-10^{-8}M_\odot/yr$, while for KT Eri, $\dot{m}<2\times10^{-10}M_\odot/yr$. V2491 Cyg shows signatures of a magnetized WD, specifically of an intermediate polar. A periodicity of $\sim$39 minutes, detected in outburst, was still measured and is likely due to WD rotation. EY Cyg is accreting at $\dot{m}\sim1.8\times10^{-11}M_\odot/yr$, one magnitude lower than KT Eri, consistently with its U Gem outburst behavior and its quiescent UV flux. The X-rays are modulated with the orbital period, despite the system's low inclination, probably due to the X-ray flux of the secondary. A period of $\sim$81 minutes is also detected, suggesting that it may also be an intermediate polar. V794 Aql had low X-ray luminosity during an optically high state, about the same level as in a recent optically low state. Thus, we find no clear correlation between optical and X-ray luminosity: the accretion rate seems unstable and variable. The very hard X-ray spectrum indicates a massive WD.
\end{abstract}

\begin{keywords}
X-rays: stars, stars: cataclysmic variables, novae: individual: V2491 Cyg, KT Eri, EY Cyg, V794 Aql
\end{keywords}

%%%%%%%%%%%%%%%%%%%%%%%%%%%%%%%%%%%%%%%%%%%%%%%%%%

%%%%%%%%%%%%%%%%% BODY OF PAPER %%%%%%%%%%%%%%%%%%

\section{Introduction}
We examined quiescence X-ray observations of V2491 Cyg
and KT Eri, two of the most luminous novae of the last 12 years,  and 
compared them with EY Cyg, a dwarf nova that is also an old nova candidate,
 likely to have had an outburst hundreds of years ago, 
 and with the VY Scl system V794 Aql, a nova-like that is 
 thought to be accreting at a high rate, yet does not
 seem to ever undergo nova eruptions. 
 We note that the possibility that VY Scl systems do have nova outbursts at some
 point has recently become very interesting after one of them has been
 observed in a nova outburst \citep{Li2020}.  

 Preliminary results of this work were published in \citet{Sun2020}.
 The theoretical models of
 the evolutionary cycle of classical and recurrent novae (CN, RN) 
 are mostly determined by two fundamental parameters: the white dwarf (WD)
 mass and the mass transfer rate $\dot m$ onto the WD \citep{Starrfield2012, Wolf2013}.
 For simplicity, most numerical models of novae outbursts
 involve a constant $\dot m$; however, there is no reason for which
 $\dot m$ should not vary. The so-called ``hibernation model''
 predicts short periods of high $\dot m$ before and after nova outbursts,
 when the nuclear burning WD irradiates the secondary so that mass transfer is enhanced \citep[see][the latter an updated version of the model]{Shara1986, Hillman2020}. $\dot m$ may also vary less regularly,
 due to interactions with the secondary \citep{Shaviv2017}. 

The observations analyzed here were part of a long term project 
 of following novae with well-known outburst characteristics and parameters,
 as they settle into quiescent
 accretion. The most recent previous articles were part of Polina Zemko's
 thesis at the University of Padova \citep{Zemko2015, Zemko2016}, and were preceded by several other 
 observations with X-ray telescopes by Orio, her group, and collaborators
 \citep[][]{Orio1993, Balman1995,
Orio1996, Mukai2005, Orio2009, Nelson2011, Mason2013, Zemko2015,
 Zemko2016, Zemko2018}. The X-ray range is particularly interesting
 to study quiescent novae and all other cataclysmic variables (CVs).
 In disk accretors, by
converting gravitational into electromagnetic energy, $\dot m$ powers the X-ray luminosity of the disk boundary layer. Thus, $\dot m$ can be estimated with a model fit,
 of plasma in collisional ionization equilibrium, often with a gradient
 of plasma temperature
 \citep[see][]{Mukai2003}. Measuring and fitting the complete X-ray spectrum is fundamental for this aim.
The hot plasma's maximum temperature is also directly related to the WD mass because the temperature of the shock is directly proportional to the potential well
of the WD. In magnetic systems, where accretion
 is funneled to the magnetic poles of the WD, a strong standing shock is formed near the
 WD surface. Thus, the X-ray flux and spectrum trace the accretion process and its rate, which gives an estimate of the WD mass. 
By and large, the X-ray emission of 
CVs correlates well with theoretical expectations \citep[e.g.][]{Pandel2005}.

Of course
 even at quiescence, CVs are complex systems with several
 components that emit in different wavelength ranges: the 
 accreting WD, the mass-loosing main-sequence (or slightly 
 evolved) secondary, the accretion disk and/or the accretion stream 
 to the poles (for highly magnetized WDs) and winds or
 other outflows from the binary. {Therefore,} the physics inferred
 from the X-ray observation varies from one system to another.
 At high accretion rates, the boundary layer may not
 remain optically thin and the 
 the  accretion disk may be flared, forming a corona
 that emits mainly at far UV wavelengths rather than at the expected X-rays 
wavelength \citep{Mauche1998}. Moreover, outflows
from the disk may also contribute to the X-rays \citep[e.g.][]{Nixon2020}. 
 
In addition to the possibility of estimating $\dot m$, its
 variations, and the WD mass, one particularly interesting problem 
 onto which X-ray observations  offer a window 
 is the effect of the WD magnetic field on nova outbursts
 and evolution. The magnetic field is not a simple variable to take into
 account in nova models, but it may be a fundamental one.
 We have found evidence that some novae are intermediate polars 
 \citep[Ps, see][]{Zemko2016, Aydi2018, Zemko2018}. 
 IPs host WDs whose magnetic field strengths reach several $10^5$ 
Gauss. An accretion disk forms, but it is
truncated where the ram pressure of the matter in the disk is equal to the
magnetic pressure of the WD's magnetic field (at the the Alfv\`en radius)
 and the matter is channeled to the poles.
WDs in IPs are not synchronized with the orbital
 period like  the more magnetized
  polars, and since the magnetic axes of WDs
are generally inclined with respect to their rotation axes, the asynchronous primary is an oblique rotator. X-ray
flux modulation with the WD spin period is one of the main observational
properties of IPs. 
Modulations due to this WD spin,
with a period of tens of minutes are typically detected in
 the X-ray light curves of IP.
 Measuring the WD spin in X-rays is considered the best indirect proof of
 the IP nature; in fact, given the range of magnetic field (10$^5$-10$^6$ Gauss)
 and the distances of known IPs, in many cases direct measurements of circular polarization only yield upper limits for the magnetic field. 
 At distances of a few Kpc, the indirect measurement
 is very often the only feasible one \citep[see][for a review]{Ferrario2015}. 

\section{Targets of these observations}
 The first three observations were done with {\sl XMM-Newton}, and we present here the broadband X-ray spectra taken with EPIC.
 The Reflection Grating Spectrographs, or RGS, did not collect enough photons;
 we obtained some optical and ultraviolet photometric measurements
 with the Optical Monitor, that we do not present
 here. The range of EPIC pn is 0.15-12 keV, but we had a large background in the softest band,
 so we limited the spectral analysis to above 0.2 keV.  There is also 
 an intense complex of detector-intrinsic lines due to Cu-K$\alpha$,
 Ni-K$\alpha $ and Z-K$\alpha $ lines around 8 keV, so 
 we preferred to fit the pn spectrum below 8.0 keV for V2491 Cyg. 
The MOS's are calibrated at 0.3-12 keV. {For both cameras, due to the very low counts above 10 keV, we extracted the data with 10 keV as upper limits.} The 0.2-10 keV count rates we detected with the
 EPIC cameras and the observation dates and exposure times are in Table 1.
 We could not use all the exposure time due to high background intervals, that we removed to obtain ``clean'' spectra. KT Eri was also observed with the ACIS-S camera of {\sl
 Chandra}. The last observation, V794 Aql, 
 was done with Chandra, the ACIS-S camera, and the High Energy Transmission
 Gratings (HETG). 
\begin{table*}
\centering
\caption{Details of the {\sl XMM-Newton} EPIC observations. The count rates
  are corrected for the background, and are in the 0.2-10 keV range
 for the pn and 0.3-10 keV for the MOS. The exposure length is the nominal exposure (individual MOS or pn exposures may be shorter by up to 2 ks. The time used
 for EY Cyg refers only to the spectral analysis, while for the timing analysis
 we could use the whole interval.}
\begin{tabular}{|c|c|c|c|}\hline
Target Name           & V2491 Cyg & EY Cyg & KT Eri \\ \hline
Observation Date (UT) & 2017 November 27 14:49:57 & 2007 April 23 12:43:52 & 2018 February 15 21:47:05 \\ \hline
Exposure Length (ks)  & 34 & 45.4 & 72.3 \\ \hline
Time used (ks)        & 22.2   & 20.0  & 51.1      \\ \hline 
MOS1 Net Count Rate (s$^{-1}$) & 0.0793$_{-0.0022}^{+0.0022}$ & 0.0941$_{-0.0028}^{+0.0028}$ & 0.0142$_{-0.0075}^{+0.0075}$ \\ \hline
MOS2 Net Count Rate (s$^{-1}$) & 0.0840$_{-0.0022}^{+0.0022}$ & 0.0995$_{-0.0032}^{+0.0032}$ & 0.0143$_{-0.0075}^{+0.0075}$ \\ \hline
pn Net Count Rate (s$^{-1}$) & 0.3168$_{-0.0050}^{+0.0050}$ & 0.3385$_{-0.0055}^{+0.0055}$ & 0.0510$_{-0.0014}^{+0.0014}$ \\ \hline
\end{tabular}
\end{table*}

\subsection{V2491 Cyg}
V2491 Cyg had a luminous nova outburst in 2008 April, reaching V=7.7 \citep{Nakano2008}, with an amplitude of 9.4-mag in the V band \citep{Jurdana2008}.
It was a fast nova, with t$_2$ (time for a decrease by 2-mag in optical)
 of 4.6 days, and the ejecta velocity reached 4860 km s$^{-1}$.
A rebrightening occurred at the end of April, then the luminosity declined to
 V$\simeq$16 after about 150 days (1 mag brighter than pre-outburst
 level). The orbital period is not known, although a modulation
 with a period of $\simeq$2.3 hours detected in outburst \citep{Baklanov2008}
 was proposed to be orbital in nature. No periodicity, however, was 
detected in quiescence by \citet{Shugarov2010}. The distance to the nova is not
 known, because the {\sl Gaia} parallax was 
not determined,  although estimates were proposed, 10.5 kpc \citep{Helton2008} and 14 kpc
 \citep{Munari2011}, based on the Maximum Magnitude versus Rate of Decline
(MMRD) relation. 
 The optical spectra of V2491 Cyg are similar to two known RN,
U Sco and V394 Cra \citep{Tomov2008},
 although
 there is no evidence, so far, of a previous outburst of this nova.
 
 In the outburst, V2491 Cyg became a very luminous supersoft X-ray source (SSS)
when
 the central source contracted, and the ejecta became transparent to supersoft
 X-rays, between day 36 and 42 \citep{Page2010}. The SSS flux increased 
 until day 57 and then started decreasing. As the X-ray flux decreased, 
 the temperature seemed to remain constant \citep{Page2010},
 indicating a possible shrinking
 of the emitting region rather than
 cooling. This resembles the cases of V407 Lup \citep{Aydi2018} and
 V4743 Sgr \citep{Zemko2016}, two novae that are thought to be IPs,
 in which the nuclear burning seemed to last longer in a {restricted}
region of the WD {surface},
 presumably at the polar caps where accretion started again.  
 The peak atmospheric temperature was very high compared with most novae,
 probably close to a million K \citep{Ness2011}, indicating 
 a massive WD.
 In this phase, V2491 Cyg showed irregular
 variability, but also a regular modulation with a 37.2-minute period
 \citep{Ness2011}. 

 V2491 Cyg is a rare case of a nova known as an X-ray source before the
 outburst \citep{Ibarra2009}, and for this reason, it was observed again in quiescence with
 {\sl Suzaku} by \citet{Zemko2015}. 
 A modulation with a period of 38 minutes (close to the period of the SSS after the outburst) was detected again, but its stability was not assessed. \citet{Zemko2015} speculated that the {detected period represented the spin 
 of an IP, and they} also found that the broadband
 X-ray spectrum was consistent with a class called ``soft IPs''.
 However, the very soft portion of the spectrum observed for V2491 Sgr was not 
clearly due to accretion. The still relatively large flux and supersoft spectrum indicated the possibility that nuclear burning in a restricted region around the polar caps was observed, either because it was quenched later in hot, accreting region, or because it was kept alive by newly accreted material. This is consistent with the interpretation of an emitting region still shrinking rather than cooling, as mentioned above, {in} the post-outburst phase.
 \citet{Zemko2018} performed optical photometry,
discovering a 36.24-minute period, slightly shorter than the one detected in X-ray. This may be the beat of the rotational and orbital period, giving higher credibility {to the hypothesis} that V2491 Cyg is also an IP.
If {the beat period is indeed the one measured in the optical
 light curve}, the orbital period has to be
of $\sim17$ hours (longer than the continuous observations that were possible during the nights in that study).

\subsection{KT Eri}
KT Eri is another nova for which we may connect a well-studied
outburst behavior with accretion and quiescence properties.
It was discovered by \citet{Yamaoka2009} at V=8.1, but probably not on the actual outburst day. \citet{Drake2009} suggested the outburst occurred after 2009 November 10.
 \citet{Hounsell2010}, a pre-discovery study, suggested that the maximum was on 2009 November 14.67 at V$\sim5.42$,
 with a pre-outburst rise of by 3 magnitudes in 1.6 days.
The outburst amplitude was 9 mag and the estimated time for
 a decay by 2 magnitudes was $t_2=6.6$ days \citep{Hounsell2010},
 indicating that KT Eri was a fast nova. The distance 
 is $3.69_{-0.43}^{+0.53}$ kpc,
 obtained thanks to the {\sl Gaia} parallax \footnote{From the Gaia database using ARI's 
Gaia Services, see  \url{https://www2.mpia-hd.mpg.de/homes/calj/gdr2_distances/gdr2_distances.pdf}\label{fn:gaia}}. 
The ejecta velocity reached a maximum of
2800$\pm$200 km s$^{-1}$ \citep{Ribeiro2011}.
\citet{Jurdana2012} found that at the nova at quiescence {varies 
 with} two distinct periods of 376 and 
737 days, and a 1 mag amplitude. These authors noted the similarity of the optical and X-ray light curve of KT Eri with those of known RNe, 
with which the nova also had a large ejection velocity in common, but no previous outburst could be found in the Harvard plates.
 \citet{Jurdana2012} thus suggested that the outbursts recur perhaps on timescales of a few centuries.
 The companion
is very likely to be an evolved star, ascending towards the red giant branch.
In X-rays, KT Eri was first detected with the {\sl Swift} XRT
as an SSS on day 39.8 after its outburst \citep{Bode2010}.
 Later, the SSS flux extremely softened on day 65.6.
 The timing analysis of the
 {\sl Swift XRT} data from day 66.60 to 79.25 revealed a 35 s
 modulation  \citep{Beardmore2010}, that was later confirmed by \citet{Ness2015}
 with {\sl Chandra}.
 \citet{Ness2015} showed that this modulation was detected only during 
the early SSS phase, and then again much later, on day 159.
With the {\sl Chandra} low energy grating spectrograph (LETG),
 the nova spectrum appeared similar to N SMC 2016 \citep{Orio2018},
 with the luminous continuum and deep absorption lines that are typical of
 many novae X-ray spectra \citep{Ness2010}.

\subsection{EY Cyg}
EY Cyg is a dwarf nova of the U Gem type, and
 it is of particular interest to us because it is  also a candidate old nova: 
 a diffuse remnant was observed in an optical image by \citet{Sion2004}. 
 The outburst may have happened about 200 years ago, according to the above authors.  The DN outbursts may indicate that it is currently transferring mass at a lower rate than at
the time it had the nova outburst \citep[see][]{Shara1986, Sion2004}.
 EY Cyg is known as a disk accretor with parameters well constrained by
optical and UV observations. \citep{Sion2004} inferred a massive
 white dwarf (WD) of 1.26 M$\odot$ and a secondary of
0.59 M$_\odot$, while \citet{Echevarria2007} estimated 
a  WD mass of $1.10\pm0.09$ M$_\odot$ and a secondary of $0.44\pm0.02$
 M$_\odot$. 
The inclination angle should be small, and \citet{Echevarria2007} constrained to
 between 13$^{\rm o}$ and 15$^{\rm o}$.

The {\sl Gaia} parallax revealed a distance of $636\pm8$ pc$^{\ref{fn:gaia}}$. 
 \citet{Connon1997} found that the secondary star is in the spectral type range
 be K5-M0, in  agreement with other authors, but
 \citet{Echevarria2007} measured the calcium lines and claimed that
this classification should be revised to K0 or late G-type.
 Sion et al. (2004) modeled the accretion disk
 with  $\dot{m}\sim10^{-10}$ M$_\odot$ yr$^{-1}$ in order to fit the far-UV spectra,
 but this model places the source at a larger distance than the Gaia
 distance, by a factor of at least 1.4,
 implying that the adopted mass accretion rate is, most likely, only an upper limit.
\citet{Echevarria2007} measured 
the orbital period of EY Cyg,   11.0237976 hours; the V magnitudes of the system during quiescence and its DN outbursts are $\sim14.8$ and $\sim11$ respectively, and the long-term AAVSO light curve revealed that the
 recurrence time of the DN outbursts is about 2000 days \citep{Tovmassian2002}. 

EY Cyg was detected for the first time in X-rays
 with ROSAT by \citet{Orio1992} in the 0.2-2.2 keV energy range.
 If it is a cooling CN a few hundred years after the outburst,
 it is interesting to compare it with
 CN a few years only after the outburst.

\subsection{V794 Aql}
V794 Aql is a VY Scl type nova-like system \citep[see][]{Zemko2014}, 
 studied by \citet{Godon2007}), who inferred a white dwarf mass of 0.9 M$_\odot$. 
The distance obtained with the {\sl Gaia} parallax is 641$\pm$21 pc$^{\ref{fn:gaia}}$.
 VY Scl binaries have occasional ``low states'' of lower optical
 and UV luminosity,  during which mass transfer seems to
 be halted \citep[see][and references therein]{Zemko2014}.
From an optical luminosity level between magnitude 14 and 15,
 V794 Aql plunges to magnitude between 18 and 20 in the low optical state.
 The accepted scenario
 is that all VY Scl in their more common high state must have high mass accretion rate,
 of the order of $10^{-8}$ M$_\odot$ yr$^{-1}$, and \citet{Godon2007}
 found indeed $\dot m=10^{-8.5}$--$10^{-8}$ M$_\odot$ yr$^{-1}$.
 We note that \citet{Zemko2014}
 estimated only a few 10$^{-11}$ M$_\odot$ yr$^{-1}$ for the VY Scl
 TT Ari in the high state. Perhaps $\dot m$ is variable even during the high states: if
 it is constantly high, it
 is not understood why it does not lead to thermonuclear burning 
\citep{Zemko2014}. 
 Contrarily to an initial hypothesis of supersoft
 X-ray luminosity raising for nuclear burning during
 the low states, \citet{Zemko2014} found instead that the X-ray
 luminosity decreases with the optically low states in V794 Aql and 
 in other VY Scl systems. In all examined cases,
 there was no luminous supersoft X-ray source witnessing nuclear burning
 during the low states.
 Zemko's 2017 Ph.D. thesis also showed that the lack of
 evidence of nuclear burning can either be explained with
 a WD whose mass is significantly lower than the Chandrasekhar mass, and/or by
low time-averaged accretion rate.  With either of these characteristics,
VY Scl are not type Ia supernova candidate progenitors,
 as previously proposed, and significantly differ from RNe, many of which seem to
 have a mass accretion rate of the order of $10^{-8}$  M$_\odot$ yr$^{-1}$. 
 The RNe are not a numerous group, and several of them are symbiotics rather than CVs. In symbiotics, it is much more challenging to trace accretion
 via X-rays observations, for several reasons, including more absorbing material
 in the symbiotic nebula, and the fact that the Roche lobe is often
 only partially filled.
 
\section{Spectral and Timing Analysis of the Observations}

\subsection{V2491 Cyg:  spectral analysis}
A 34 ks observation of V2491 Cyg with {\sl XMM-Newton} (Observation ID: 084580101) was
 proposed by PI Orio and was done on 2017 November 27,
 3519 days after the observed optical maximum. We obtained 22.2 ks of good exposure time, not affected by elevated background flares.
 The observations were done with the thin filter; the pn was
in large window mode, the MOS1 in partial large window mode (``prime W3''),
 the MOS2 in large window mode.
 The RGS spectra did not have sufficient signal-to-noise; we only
 analyze the EPIC X-ray data here.

Fig. 1 shows the comparison of the {\sl Suzaku} spectrum observed
 on November 20 2011. The EPIC pn is sensitive at lower energy 
 than {\sl Suzaku} and has a large effective area at very soft energy,
 while {\sl Suzaku} is more sensitive than EPIC around 6.5-7 keV, where the
 main iron lines are observed. Despite these differences, we established 
 that the 2011 spectrum was somewhat harder. Even if the iron lines are
 not well resolved with EPIC, it seems the iron reflection line at 6.4 keV of neutral iron was less intense in 2017 because, in the EPIC spectra,
the iron complex seems to had been dominated by the H-like and He-like lines (Fe XXVI and Fe XXV). 

Two different models fit the XMM-Newton spectra of V2491 Cyg. Only one spectral component
 does not fit the whole spectrum; 
the less complex  model we could use (Model 1 in Table 2) includes two absorbed 
components: a blackbody and a  thermal plasma in collisional ionization equilibrium,
calculated with the Astrophysical Plasma Emission Code (APEC) in XSPEC \citep{Smith2001}. 
 We modeled the interstellar absorption with the T\"ubingen-Boulder 
TBABS model \citep{Wilms2000}.
 The plasma temperature is not constrained in this fit and within
 90\% probability it may be as high
as the maximum value (64 keV) of the model; so we fixed it at 10 keV, close
to the {\sl Suzaku} temperature, to constrain at least the other parameters.
 The second model (Model 2 in Table 2) is the one that fitted  
 the {\sl Suzaku} data in work by \citet{Zemko2015}; it includes a second thermal component at a very low temperature,
 and for the hotter plasma, a partially covering absorber, which is often used 
 to fit IP spectra, because of the accretion column partially obscuring the emission.
 This model also includes a Gaussian to fit the iron complex. Although
 there is a need of a separate line for the reflection line of iron at 6.4 keV, it is not clearly resolved from the iron Fe XXV triplet around 6.7 keV that,
 in the XMM data, it appears stronger than the line at 6.4 keV.
 Perhaps this is due to enhanced iron with respect to the solar value,
 but we did not have enough signal to noise and resolution to assess this. 
 The absorption is modeled  with WABS, the Wisconsin ``recipe''
\citep{Morrison1983}, not with TBABS, as in our models in this paper.
The fit parameters are in Table \ref{tab:v2491spec}.
 The EPIC-pn spectrum and the fit with Model 2 are shown in Fig. 1 
 together, for comparison, with the Suzaku spectrum and fit by
 \citet{Zemko2015}; the
three EPIC spectra fitted with Model 1 are shown in Figure \ref{fig:v2491spec}. 
\begin{figure}
\centering
\includegraphics[width=0.5\textwidth]{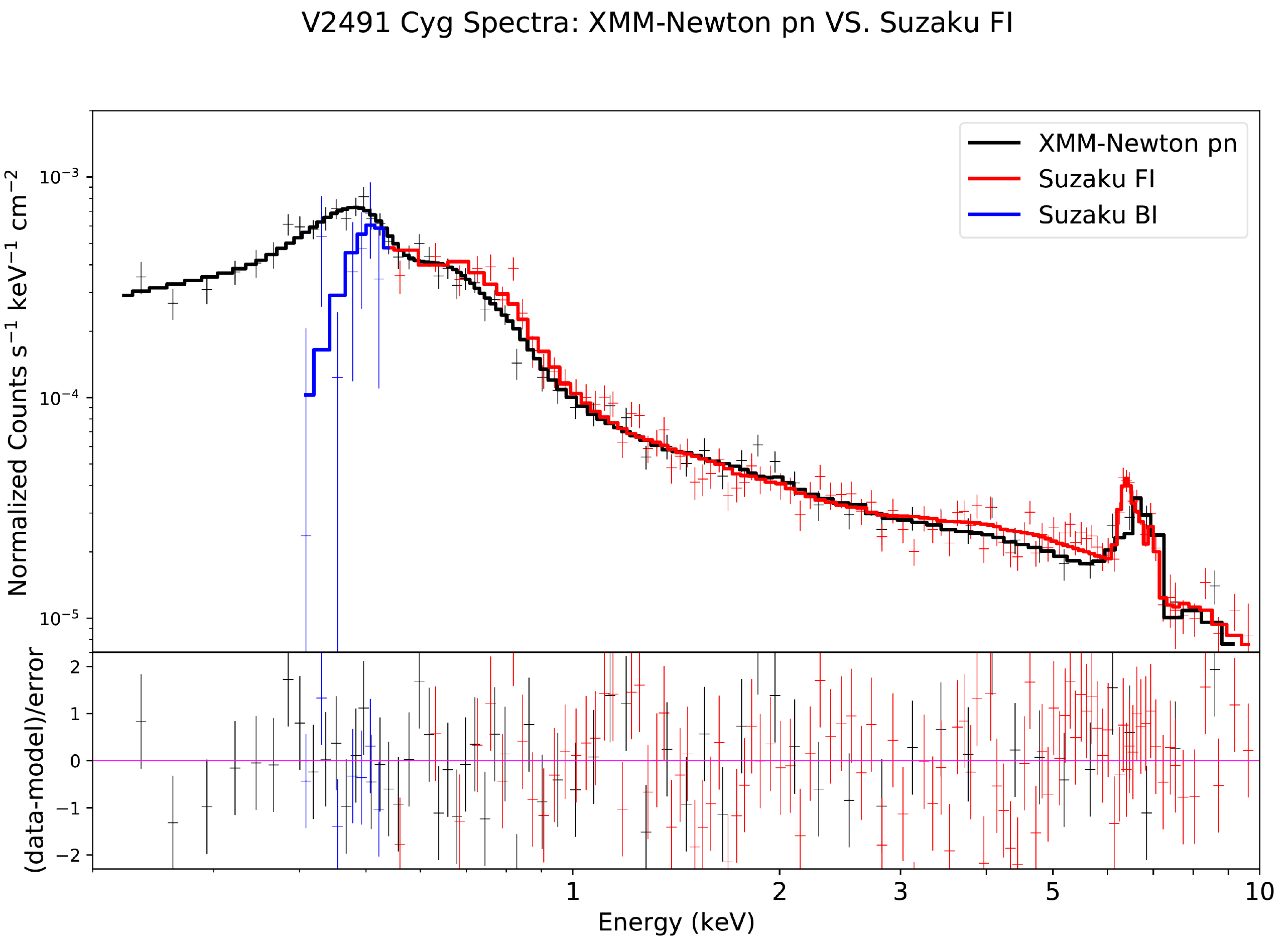}
\caption{\label{fig:v2491spec} V2491 Cyg EPIC pn spectrum in black
 and Suzaku FI (front illuminated) CCD spectrum (in red) and
 the soft portion of the BI (back illuminated) spectrum in blue. 
The best fit with the model in the second
 and third column of Table 1 (the model of \citet{Zemko2015}
 for the {\it Suzaku} spectrum and its application to the EPIC pn
 one,  are also shown.
}
\end{figure}
\begin{figure}
\centering
\includegraphics[width=0.5\textwidth]{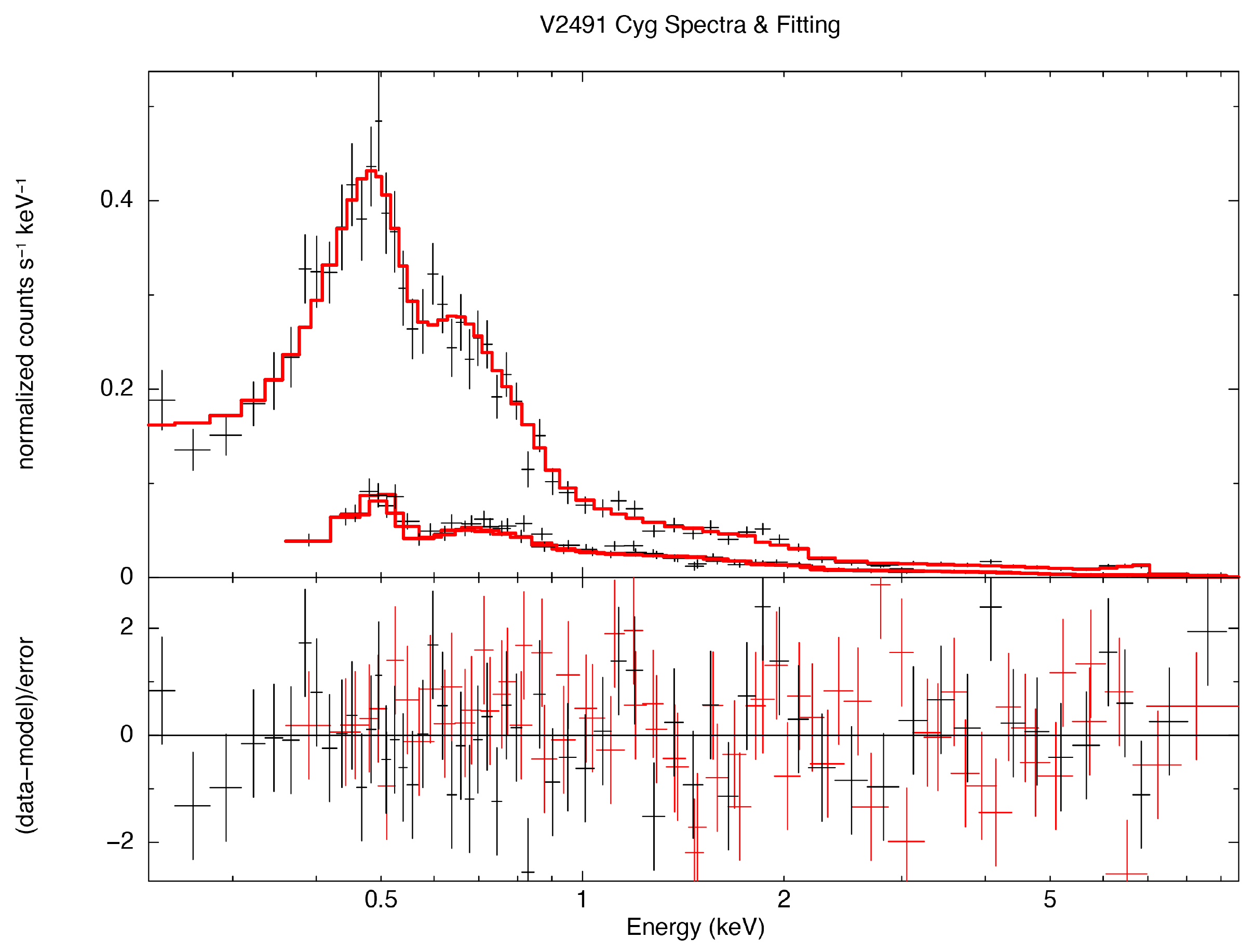}
\caption{\label{fig:v2491spec} V2491 Cyg EPIC spectra in units
 of count rate. The pn data are
 the higher ones, MOS are the lower values.
The best fit with Model 2 is in red.
 The residuals for the pn are in black and for the MOS in {red}.
}
\end{figure}
\begin{table*}
\centering
\caption{
\label{tab:v2491spec}
Best fit parameters for the V2491 Cyg combined
 EPIC spectra with Model 1 (using TBABS) and Model 2 (using WABS, and
 the plasma temperature of
 the best fit by the above authors for Suzaku).
$L_{bb}$ is the bolometric luminosity of the blackbody, assuming a distance of 10.5 kpc.
 $F_{\rm abs}$ and $F_{\rm unabs}$ the absorbed and unabsorbed X-ray flux, in
 different energy ranges.
Errors are estimated at the 90\% confidence level.
}
\begin{tabular}{|c|c|c|c|}\hline
Parameter/Result & Model 1 & Model 2 & Suzaku (Zemko et al. 2015) \\ \hline
N(H) (wabs, $\times10^{22}$ cm$^{-2}$) & $0.40_{-0.04}^{+0.05}$ & $0.30_{-0.02}^{+0.04}$ & $0.25_{-0.06}^{+0.08}$ \\ \hline
Blackbody Temperature (eV) & $76_{-3}^{+3}$ & $82_{-7}^{+3}$ & $77_{-9}^{+7}$ \\ \hline
$L_{bb}$($\times10^{35}$ erg s$^{-1}$) & $2.77_{-0.98}^{+1.62}$ & $1.03_{-0.37}^{+0.61}$ & $1.4_{-0.7}^{+2.4}$ \\ \hline
N(H) (pcfabs, $\times10^{22}$ cm$^{-2}$) & - & $14.9_{-4.0}^{+5.0}$ & $13.3_{-2.4}^{+3.0}$ \\ \hline
Partial Covering Fraction & - & $0.57_{-0.07}^{+0.06}$ & $0.66_{-0.03}^{+0.03}$ \\ \hline
$T_1$ (keV) & - & 0.24 (Fixed) & $0.24_{-0.24}^{+0.24}$ \\ \hline
$T_2$ (keV) & $10.0$ (Fixed) & 11.3 (Fixed) & $11.3_{-1.5}^{+1.8}$ \\ \hline
Norm$_1$ ($\times10^{-4}$ cm$^{-5}$) & - & $0.0_{-0.0}^{+2.2}$ & $5_{-5}^{+8}$ \\ \hline
Norm$_2$ ($\times10^{-4}$ cm$^{-5}$) & $5.5_{-0.6}^{+0.6}$ & $11.7_{-0.9}^{+2.0}$ & $14.6_{-1.0}^{+1.1}$ \\ \hline
Gaussian  EW  (Fixed) (eV) & 245 & 245 & 245 \\
\hline
$\chi^2$ & 1.6 & 1.1 & 1.1 \\ \hline
$F_{\rm abs,0.2-1.0keV}$ ($\times10^{-13}$ erg cm$^{-2}$ s$^{-1}$) & $2.43_{-0.34}^{+0.04}$ & $2.44_{-0.27}^{+0.37}$ & 4.38 \\ \hline
$F_{\rm abs,0.2-10.0keV}$ ($\times10^{-12}$ erg cm$^{-2}$ s$^{-1}$) & $1.40_{-0.08}^{+0.04}$ & $1.64_{-0.07}^{+0.07}$ & 2.10 \\ \hline
$F_{\rm unabs,0.2-1.0keV}$ ($\times10^{-12}$ erg cm$^{-2}$ s$^{-1}$) & 14.71 & 5.84 & 7.98 \\ \hline
$F_{\rm unabs,0.2-10.0keV}$ ($\times10^{-12}$ erg cm$^{-2}$ s$^{-1}$) & 15.98 & 7.31 & 9.71 \\ \hline
\end{tabular}
\end{table*}

The results indicate a significant change (about 44\% in the best fit) in the unabsorbed
 flux and by about 22\% in the unabsorbed flux;
this is due mostly to the flux between 0.7 and 1 keV as 
it is evident in Fig.1.  We plotted here the spectrum of the FI
 (front-illuminated) CCD above 0.8 keV, because it
  has the best signal-to-noise, but, additionally, 
 below 0.8 keV we plotted the spectrum of
 the BI (back-illuminated) CCD that is sensitive and calibrated
 at softer energy.
The blackbody component may not have varied
 much, but the {\sl Suzaku} instruments did not constrain it as well as EPIC
 did. On the other hand, the {\sl Suzaku} cameras were much more
 sensitive to the
 6.4-7.0 keV range in which the iron complex is measured, giving other
 diagnostics.
We find that the unabsorbed flux in the 0.7-1 keV range has decreased by 
 45\% since the Suzaku observation, and the unabsorbed one by up to 25\%, 
 as Fig. 1 shows. On the other hand, with {\sl XMM-Newton} we detected
 copious blackbody-like X-ray flux below 0.8 keV, corresponding
 to a luminosity of 2.77 $\times 10^{35}$ erg s$^{-1}$ when we account for the
 absorbing column and assume a 10.5 kpc distance. 

 We follow \citet{Patterson1985} and other authors, in estimating $\dot m$ 
 by assuming that about half of the gravitational energy due to accretion is converted in X-ray luminosity at the boundary layer.
 Those authors considered the unabsorbed X-ray luminosity in the 0.2-4.0 keV range and made
 the assumption that it is about a quarter of the X-ray luminosity at all wavelengths. Of course, this depends on the slope of the spectrum.
 We estimated  the unabsorbed X-ray flux in the 0.2-10 keV range
 and  since V2491 Cyg does have
 a hard tail, we extrapolated the model flux at hard energy 10-100 keV,
 but found it may represent
 only about 5\% of the unabsorbed X-ray flux in the 10-100 keV range. 
 After a simple manipulation
 the resulting formula, equivalent to rescaling Equation No.4 of the above authors,
 we obtained 
\begin{align}
\label{eq:m16}
\dot{M}_{16}=\frac{L_x(0.2-10.0 keV)}{33.0\times10^{31}M_{0.7}^{1.8}} g\ s^{-1}
\end{align}
{where $\dot{M}_{16}$ is the mass accretion rate in units
 of $10^{16}$ g s$^{-1}$}.
If we include all the X-ray flux, including the blackbody component,
the result is  $\dot m = 1.26 \times 10^{18}$ g s$^{-1}=1.97 \times 10^{-8}$ M$_\odot$ yr$^{-1}$ (assuming a 10.5 kpc distance),
 comparable with the estimate for RNe. If 
 the blackbody-like flux is not associated
 with accretion and it is still atmospheric, it contributes to 
a large fraction of the total flux. In fact, 
 the bolometric luminosity of the blackbody model is of the same
 order of magnitude as the total X-ray flux for a  
 10.5 kpc distance.  As hypothesized and suggested
 by \citet{Zemko2015}, we may instead be
still observing a very hot stellar atmosphere, which is about
 10 times less luminous than a blackbody at the same temperature
 (see the alternative fit with the atmosphere in that paper). In this case,
 we need to exclude this large, soft portion of the X-ray flux in accounting for
 luminosity due to accretion. For the sake of simplicity, 
 we excluded all unabsorbed flux below 1 keV (this includes more
 than the blackbody-like component's contribution) and found that the
 unabsorbed X-ray flux is
1.36 $\times 10^{-12}$ erg cm$^{-2}$ s$^{1}$, which indicates $\dot m$ of order 10$^{-9}$ M$_\odot$ yr$^{-1}$ at 10.5 kpc distance.
 We thus estimate the $\dot m$ value to be between this and the above value.
 We note that $\dot m$ scales with the square of the distance ratio,
 so at 14 kpc the range would increase by a factor of 1.78. 
\subsection{V2491 Cyg: timing analysis}
\begin{figure*}
\centering
\includegraphics[width=0.8\linewidth]{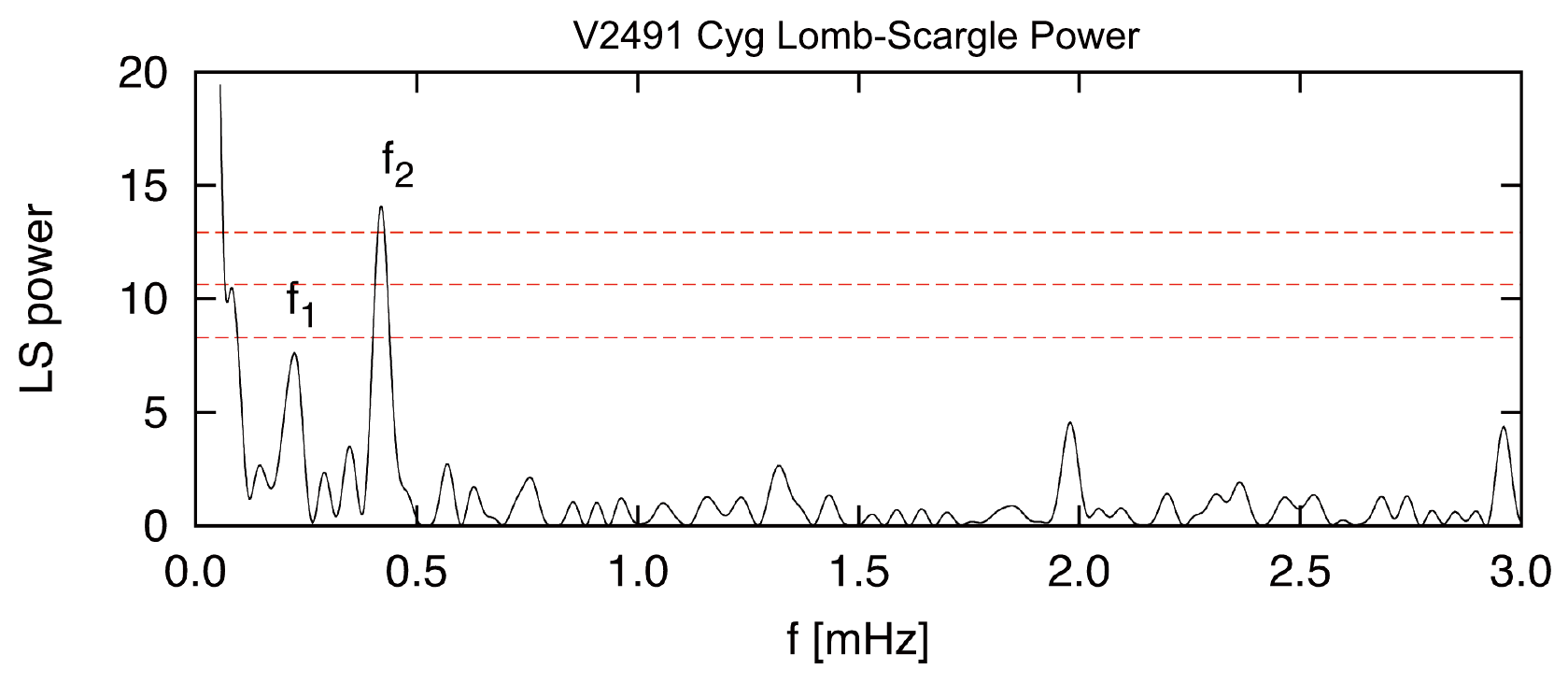}
\caption{Periodogram of non-de-trended EPIC pn light curve
 of V2491 Cyg. The red lines represent the 99.9\%, 99.0\% and 90.0\% confidence levels.}
\label{power}
\end{figure*}

\begin{figure}
\centering
\includegraphics[width=0.8\linewidth]{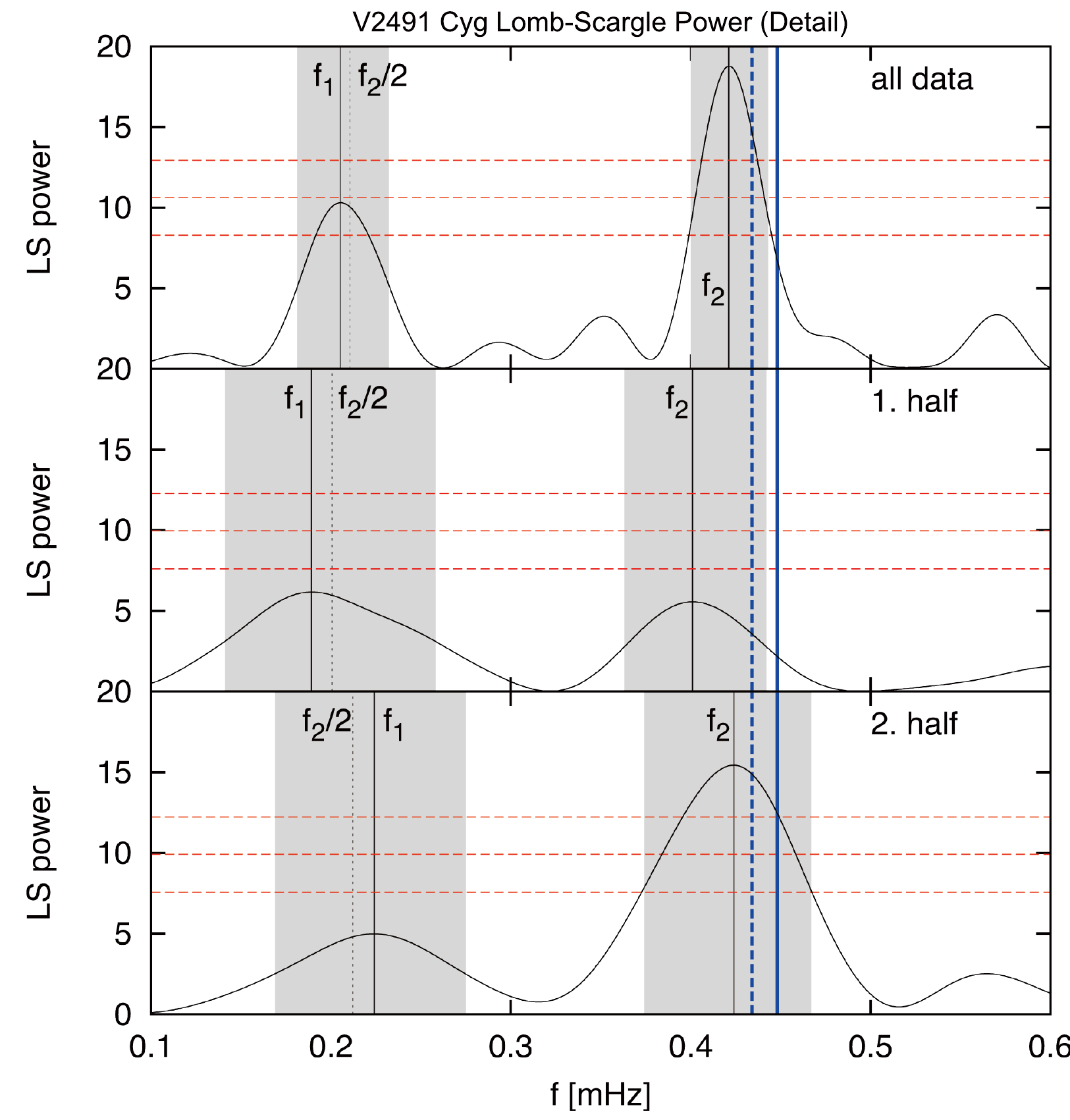}
\caption{Periodograms calculated with the de-trended EPIC pn on the light curve of V2491 Cyg.
 The vertical black solid lines represent the detected frequencies,
 with error intervals represented by shaded areas.
The  vertical, dashed black lines show the $f_2/2$ values. 
The blue lines indicate the frequencies detected in two previous observations,
 on day 40 after the outburst (solid line, \citep{Ness2011}), 
on day and 937 \citep[dashed line,][]{Zemko2015}).
The red lines indicate the 99.9\%, 99.0\% and 90.0\% confidence levels.}
\label{power_detail}
\end{figure}

\begin{figure}
\centering
\includegraphics[width=0.98\linewidth]{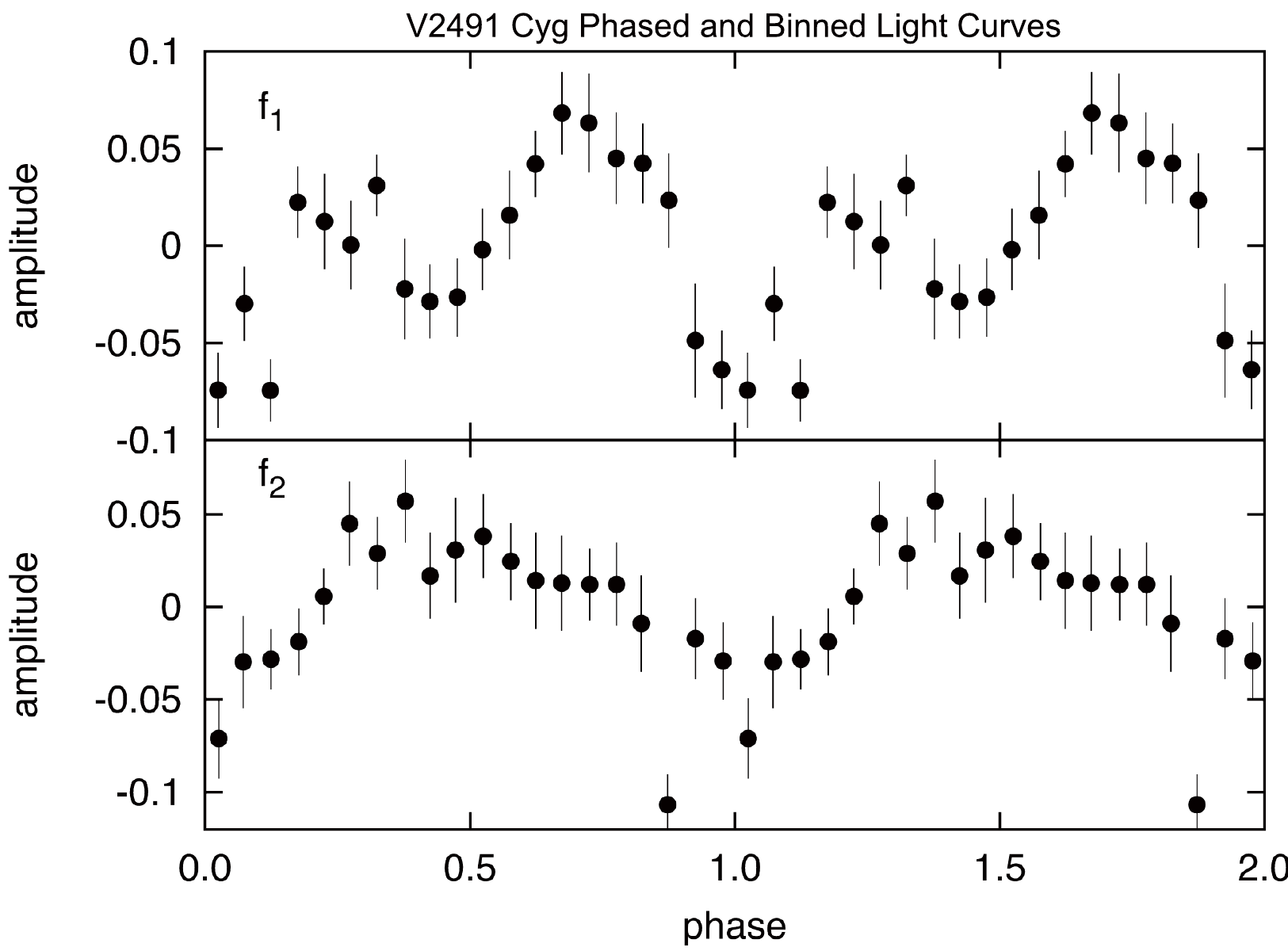}
\caption{Phased and binned whole light curves of V2491 Cyg using two different frequencies marked as label in the upper left corner. The error-bars represent errors of the mean.}
\label{lc_binned}
\end{figure}

The Lomb-Scargle periodogram (\citealt{scargle1982}) of the EPIC pn
light curve, without any de-trending, is shown in Fig.~\ref{power}. 
 Here and in the following analysis, we calculate the statistical error of the frequency as half-width at half amplitude.
 The periodogram is dominated by two peaks, $f_1$ and $f_2$. 
 The first frequency has power below the 90\% confidence range, 
while the second one has power
 above the 99.9\% confidence range. $f_2$ is the known $\simeq$38\,min
 periodicity,
 We did not detect any frequency with high power in the MOS1 light curve,
 but we did detect the frequency 
0.432$^{+0.021}_{0.020}$ mHz (corresponding to $f_2$ and to a period of 38.6 minutes)
 with a significance of 99\% in the MOS2 data. 
 This frequency value
 is within the error bar of the pn measured $f_2$.
 We will continue the discussion 
 referring to the EPIC pn, since it offers a higher S/N. 
After detrending the EPIC pn light curve with a
 4$^{\rm th}$ order polynomial, also the $f_1$ peak became 
significant, just slightly below the
 99\% confidence level (top panel of Fig.~\ref{power_detail}). 
We experimented by splitting the detrended light curve into two halves,
 and found that both peaks remained dominant in both periodograms 
(middle and bottom panel of Fig. ~\ref{power_detail}), but the significance
 for both peaks dropped considerably below the 90\% confidence
 level in the first half, while $f_2$ 
 was detected at a high confidence level in the second half.
\begin{table*}
\begin{center}
\caption{Frequencies in the
 EPIC pn and Suzaku light curves of V2491 Cyg, and days after the outburst 
 optical maximum.}
\begin{tabular}{llccc}
\hline
day & curve portion  & $f_1$ & $f_2$ & Reference\\
& & (mHz) & (mHz)\\
\hline
40 & pn, post-dip portion & & $0.448 \pm 0.036$ & \citet{Ness2011} \\
\hline
937 & pn, all & & $0.434^{+0.003}_{-0.002}$ & \citet{Zemko2014} \\
\hline
3519 & pn, all & $0.205^{+0.027}_{-0.024}$ & $0.421^{+0.022}_{-0.021}$ & this work\\
\hline
3519 & pn, first half & $0.189^{+0.069}_{-0.048}$ & $0.401^{+0.041}_{-0.038}$ & this work\\
\hline
3519 & pn, second half & $0.224^{0.051}_{0.055}$ & $0.424^{0.043}_{-0.050}$ & this work \\
\hline
3519 & MOS2, all & & 0.432$^{+0.021}+_{-0.020}$ & this work \\
\hline
\end{tabular}
\end{center}
\label{frequencies}
\end{table*}
All the values of the measured frequency are listed in Table~\ref{frequencies}.
The binned light curves are plotted
 in phase  using $f_1$ and $f_2$ in Fig.~\ref{lc_binned}. We note the double-humped
 structure in the $f_1$ plot.
 
Even if the frequencies detected in this work do not show any significant
 change after splitting the light curve into two
 equal segments, which suggests stability, 
the errors are too large for a conclusive statement on the
 stability of the modulations. However, 
first of all, we confirm the previously detected period near 38\,min. 
Second of all, the exact value derived from our {\it XMM-Newton} EPIC-pn
data is slightly longer, 39.6\,min and
Fig.~\ref{power_detail} suggests a possible continuous shift of the periodicity toward longer periods, but the measurement uncertainties in all 
three epochs do not allow us to make a definite statement.
 All the measured values at different epochs are within each other's errors.

The third important point is the following:
 the detection of $f_1$, which is very close to half $f_2$,
 implies that $f_2$ may be the first harmonic of $f_1$. 
The confidence of this pattern in the two subsamples 
(the light curve split into two halves), after de-trending,
 is below 90\%, which suggests that this pattern must be taken with caution. If it is real, the fundamental frequency may be $f_1$, while 
$f_2$ is its first harmonic. The detected periodicity of 39.6\,min would be (within the errors) half of the white dwarf rotation period. 
The frequency $f_1 = 0.205$\,mHz  yields a 
 value of the period of  81.3\,min.

\begin{figure}
\centering
\includegraphics[width=0.98\linewidth]{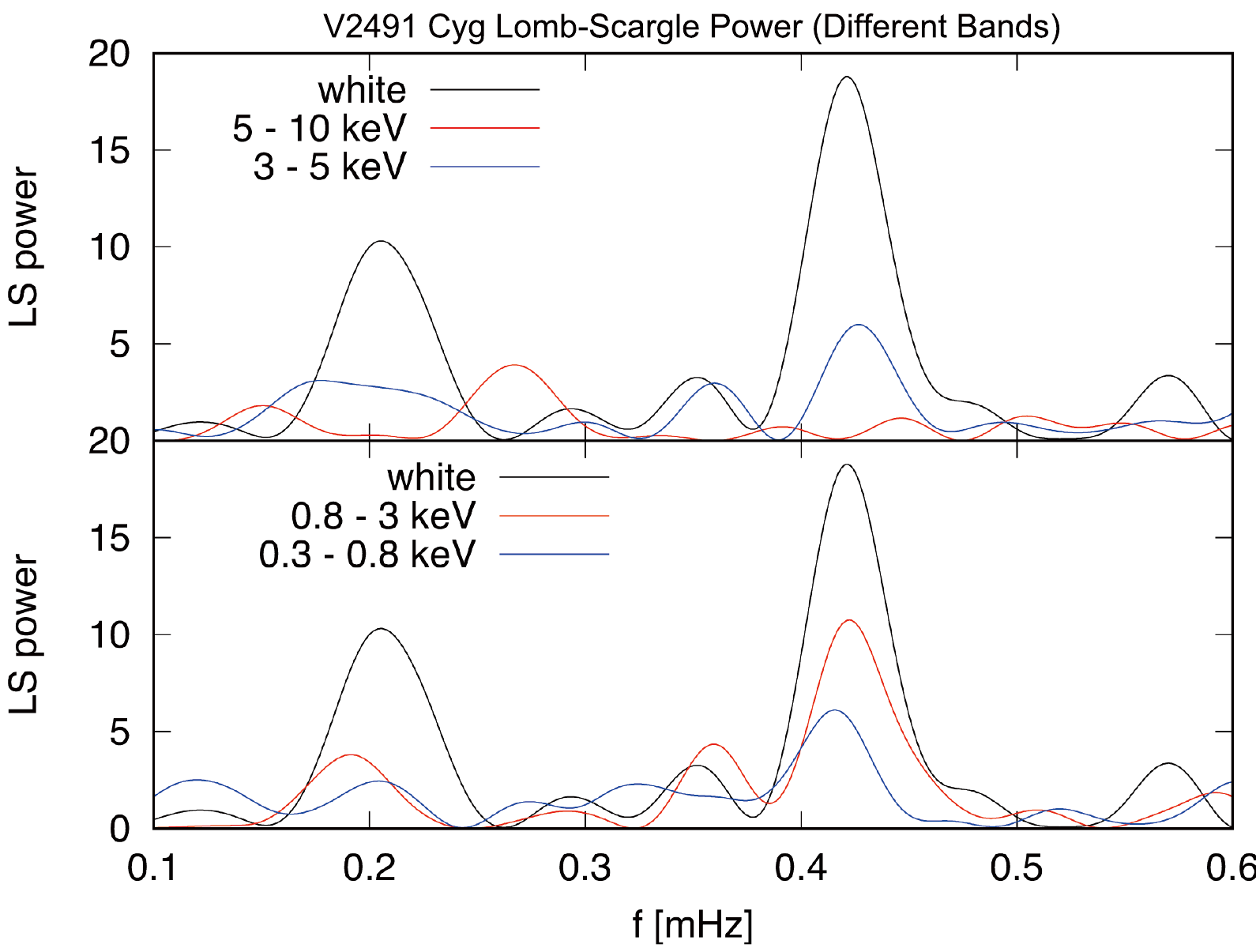}
\caption{The same as the top panel of Fig.~\ref{power_detail},
 with added periodograms obtained in different energy bands for comparison.
``White'' means 0.2-10 keV.}
\label{power_bands}
\end{figure}

In order to assess the  energy dependence of the variability, we extracted 
four light curves in different energy bands:  5-10 keV (band 1),
 3-5 keV (band 2), 0.8-3 keV (band 3) and 0.3-0.8 keV
(band 4). The count rates in each band are
 0.03 cts s$^{-1}$ in band 1 and 2,
 0.10 cts s$^{-1}$  in band 3, 0.17 cts s$^{-1}$ in band 4.
 Because of the lower count rate in each narrow band, the power of
the detected periodogram peak decreases, but
 we compared the corresponding periodograms with the 0.3-8 keV
 light curve in Fig.~\ref{power_bands}. After accounting for the different 
 average count rates, we still agree with the conclusion of \citet{Zemko2015} 
that the largest modulation is at 0.8-3 keV, although it is
almost comparable even at  3-5 keV (and is hardly measurable in band 1, that has the
 same count rate). 

\subsection{The KT Eri X-ray spectra}
When the nova returned to quiescence, we examined an archival 
Chandra ACIS-S observation
 (PI Fred Walter) done on 2011 November 13, almost exactly 3 years after
 the eruption (day 1094 after the inferred optical maximum).
The energy range is 0.3-10 keV. This exposure lasted for 75 ks 
 revealed an intriguing soft portion of the spectrum of KT Eri in quiescence, which 
resembled the first post-outburst quiescent X-ray spectra of V4743 Sgr and V2491 Cyg. 
 The Chandra spectrum and the best fit to it are shown in Fig. 7 
%\ref{fig:kterichandra}
 and in Table 4.
% \ref{tab:kterispec}. 
Although the fit is statistically acceptable,
the residuals around 0.6 keV with a solar abundance model may be due to 
 enhanced oxygen.

The nova was observed again
 by us (PI Orio) with {\sl XMM-Newton}  (Obs. ID: 0804580201)
 on February 15, 2018, on day 3378 after the inferred optical maximum. 
The observations were done with the thin filter, the pn and the MOS2 were used in full window mode and MOS1 in large window mode.
The nominal exposure of 72.3 ks was partially affected by a very elevated background, so we used 51.1 ks of good exposure time.

 Fig. 7 shows also the comparison
 between the {\sl Chandra} exposure and the {\sl XMM-Newton} one.
The spectrum became harder during the intervening years. 

\begin{figure}
\centering
\includegraphics[width=0.50\textwidth]{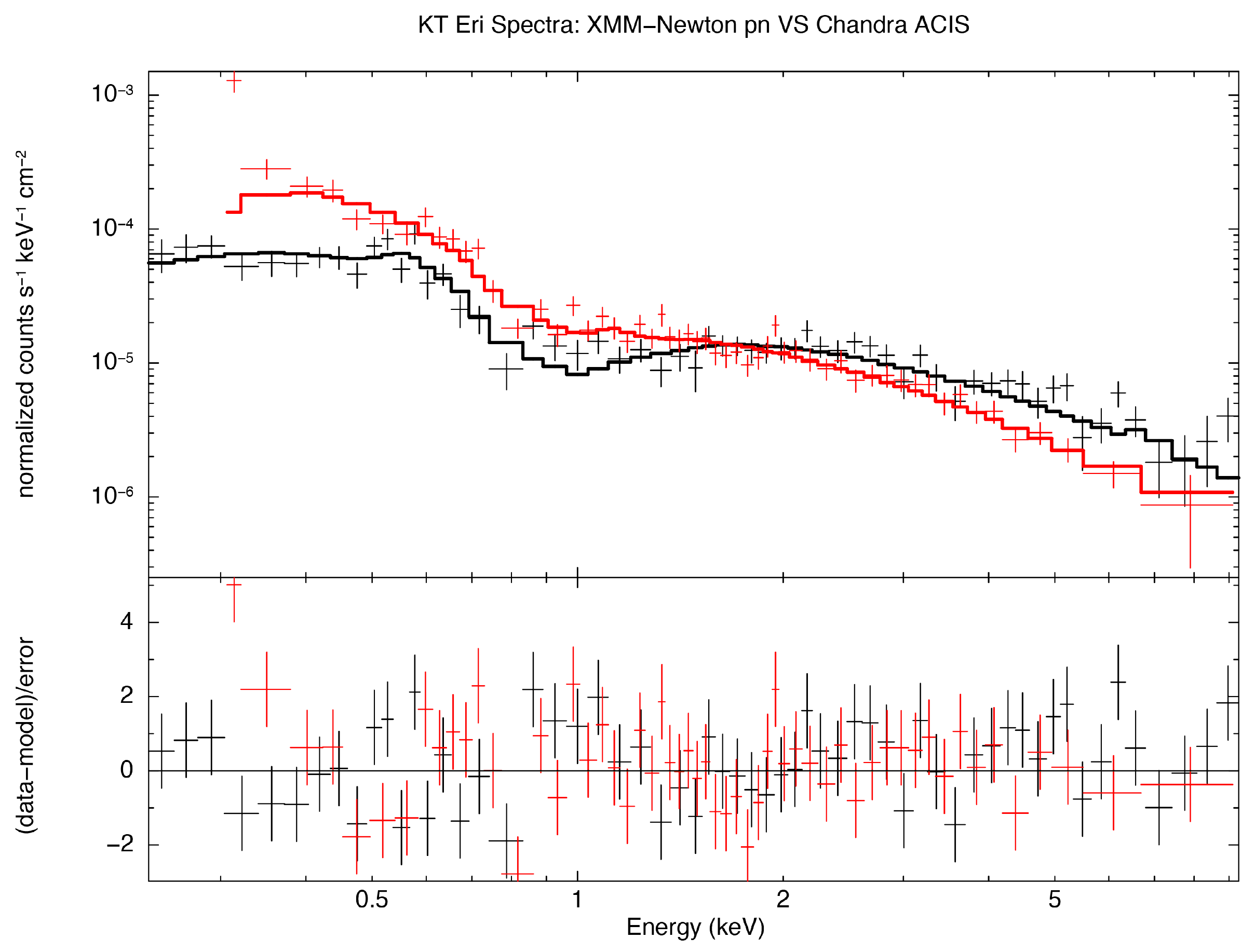}
\caption{The Chandra ACIS spectrum of KT Eri (in red, {observed in 2011 November})
 and the XMM-EPIC pn one {(observed in 2018 February)}, in black. 
 The y-axis shows the photon counts and the fits are Model 2 in Table 2,
 and the Suzaku model in the same table. }
\end{figure}
\begin{figure}
\centering
\includegraphics[width=0.49\textwidth]{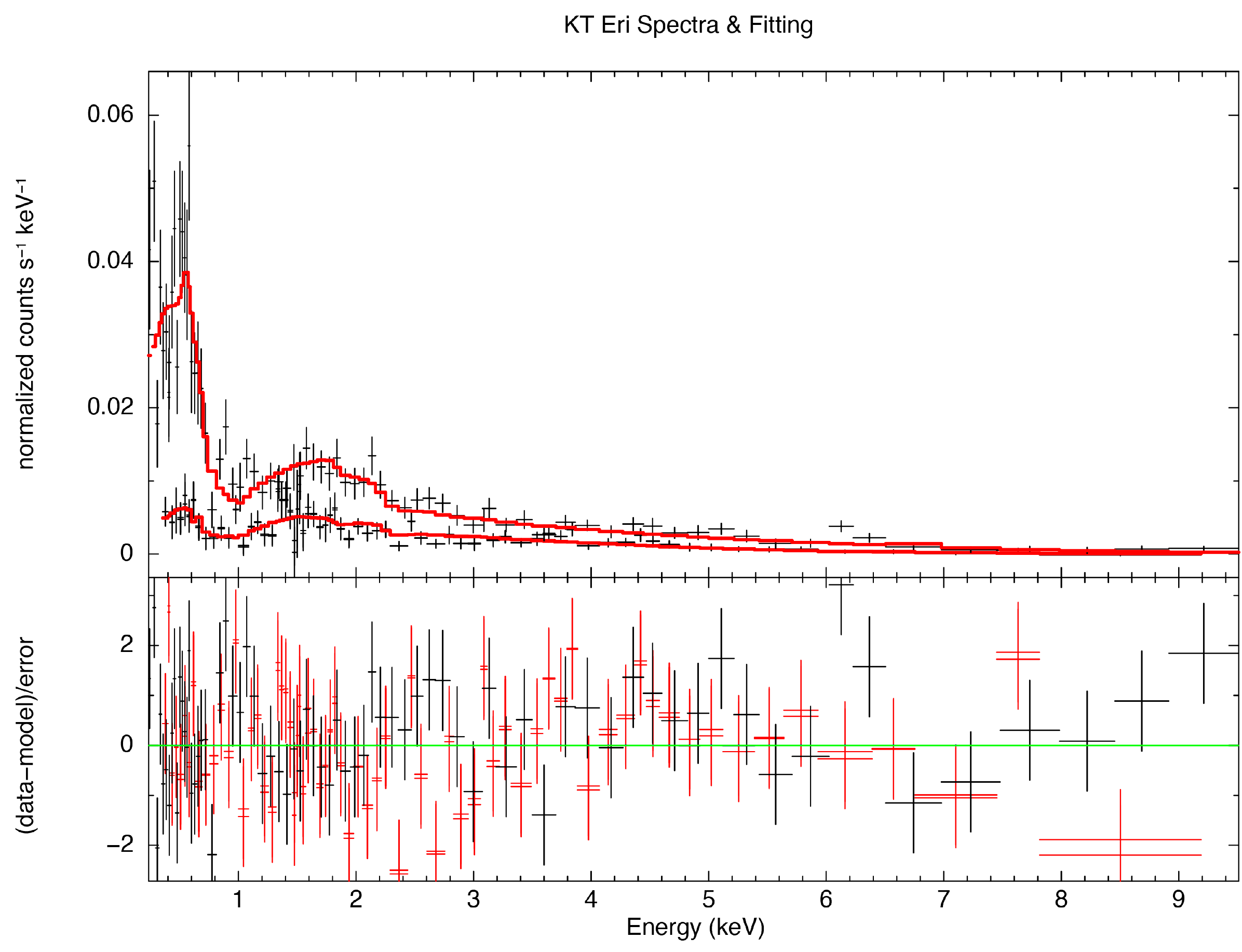}
\caption{\label{fig:kterispec} XMM-Newton spectra of KT Eri and the best
spectral fits with Model 2. The colors are the same as in Fig. 2 and
 the pn again has the higher values.}
\end{figure}
We fitted the three EPIC spectra simultaneously. 
The fit with Model 2 is shown in Figure \ref{fig:kterispec} and the fit parameters are in Table 4.
 In both the  {\sl XMM-Newton} and the {\sl Chandra} observations,
an APEC low-temperature thermal component of plasma in collisional
 ionization equilibrium  fits the softest portion of the spectrum
 a little better than a blackbody. This
 soft flux decreased by more than half between
 the time of the {\sl Chandra} and the {\sl XMM-Newton}
 observation, but the decrease {is explained by} the model as due
 to higher absorption (we can only speculate that it
 may have been intrinsic and due to wind from
 the disk). In any case, the unabsorbed luminosity of
 this component is about 2 $\times 10^{32}$ erg s$^{-1}$,
 three orders of magnitude less than for V2491 Cyg.
 Another plasma component is added
 for {\sl Chandra} in
 Model 1, while a power law is added in Model 2. We could not constrain
 the temperature of the second thermal component, which, in the XSPEC fit,
 became
 arbitrarily high, outside the EPIC range, like in
 Model 1, so we fixed at 10 keV. The power law provides an acceptable fit
and turns out to be the dominant flux component in this fit,
although a non-thermal component in a CV is not usually detected. 

 We do know that the column density to the nova is low, because of the luminous
 supersoft X-ray source detection;  an analysis of the outburst {\sl Chandra} spectra by
 Pei et al. (private communication) indicates with 90\% confidence probability that   
 N(H)$< \times 10^{21}$ cm$^{-2}$. However, we can only fit the data with column 
 density N(H)$= 10^{21}$ cm$^{-2}$ for the soft component and much higher column
 density for the hard component, because the spectrum above 1 keV is flatter 
 and quite different from other novae and CVs.
 We suggest that the system may be surrounded
 by a strong disk wind that causes intrinsic absorption.  
 We also attempted introducing a partially covering absorber like for V2491 Cyg, but 
 the resulting covering fraction would exceed 80\% and thus does not seem very
 significant when compared to much smaller values obtained
 for known IPs.

 The X-ray luminosity of KT Eri, at about 3.7 kpc distance, translates with Equation 1
 into $\dot m \simeq 1.9 \times 10^{-10}$ M$_\odot$ yr$^{-1}$, two orders of magnitude
 less than V2491 Cyg.

 The timing analysis did not reveal any significant periodicities, and we did not have high enough S/N to retrieve the 35 s modulations. Such short periods, however, are typically observed only during outburst \citep[see][]{Ness2015}.

\begin{table*}
\centering
\caption{KT Eri Spectral Fits Parameters }
\begin{tabular}{|c|c|c|c|}\hline
Parameter & 2 APECs & APEC + Power-law & Chandra 2 APECs \\ \hline
N(H)$_1$ ($\times10^{21}$ cm$^{-2}$) & 1.0$_{-0.5}^{+0.7}$ & 1.0$_{-0.5}^{+0.6}$ & 0.5 (Fixed) \\ \hline
N(H)$_2$ ($\times10^{21}$ cm$^{-2}$) & 9.4$_{-1.2}^{+1.3}$ & 6.8$_{-1.7}^{+1.9}$ & 5.6$_{-1.9}^{+2.2}$ \\ \hline
$T_1$ (keV) & 0.18$_{-0.01}^{+0.02}$ & 0.18$_{-0.03}^{+0.02}$ & 0.19$_{-0.03}^{+0.04}$ \\ \hline
$T_2$ (keV) & 10.0 (Fixed) & - & 5.18$_{-1.41}^{+2.37}$ \\ \hline
$F_{\rm 1,abs}$ ($\times10^{-13}$erg cm$^{-2}$ s$^{-1}$) & 0.25 & 0.25 & 0.64  \\ \hline
$F_{\rm 1,unabs}$ ($\times10^{-13}$erg cm$^{-2}$ s$^{-1}$) & 1.28 & 1.18 & 1.17 \\ \hline
$F_{\rm 2,abs}$ ($\times10^{-13}$erg cm$^{-2}$ s$^{-1}$) & 2.85 & - & 1.83  \\ \hline
$F_{\rm 2,unabs}$ ($\times10^{-13}$erg cm$^{-2}$ s$^{-1}$) & 4.33 & - & 2.77  \\ \hline
%$Norm_1$ ($\times10^{-4}$ cm$^{-5}$) & 6.59$_{-3.71}^{+13.65}$ & 6.07$_{-3.57}^{+11.25}$ & 12.95$_{-3.74}^{+6.04}$ \\ \hline
%$Norm_2$ ($\times10^{-4}$ cm$^{-5}$) & 2.30$_{-0.25}^{+0.25}$ & - & 1.80$_{-0.30}^{+0.37}$ \\ \hline
PhoIndex & - & 1.28$_{-0.16}^{+0.17}$ & - \\ \hline
%Norm$_{plaw}$ ($\times10^{-5}$ photons keV$^{-1}$ cm$^{-2}$ s$^{-1}$ at 1 keV) & - & 4.02$_{-0.80}^{+0.17}$ & - \\ \hline
$F_{\rm plaw,abs}$ ($\times10^{-13}$erg cm$^{-2}$ s$^{-1}$) & - & 3.39 & - \\ \hline
$F_{\rm plaw,unabs}$ ($\times10^{-13}$erg cm$^{-2}$ s$^{-1}$) & - & 4.43 & - \\ \hline
$F_{\rm abs}$ ($\times10^{-13}$erg cm$^{-2}$ s$^{-1}$) & 3.10$_{-0.21}^{+0.16}$ & 3.64$_{-0.43}^{+0.15}$ & 2.47$_{-0.26}^{+0.18}$  \\ \hline
$F_{\rm unabs}$ ($\times10^{-13}$erg cm$^{-2}$ s$^{-1}$) & 5.07 & 5.61 & 3.94  \\ \hline
$\chi^2$ & 1.2 & 1.1 & 1.8 \\ \hline
\end{tabular}
\end{table*}
\subsection{EY Cyg spectral analysis}
EY Cyg was observed on April 27 2007
 (Obs. ID: 0400670101, PI Orio); the exposure lasted for each instrument,
 pn and MOS, lasted for 43.5 ks (about 12.08 hours), 
 covering the orbital cycle of about 11 hours about
 1.1 times. 
The observations were done with the thin filter,  the pn and the two MOS
 were all in full-frame mode.

\begin{figure}
\centering
\includegraphics[width=0.50\textwidth]{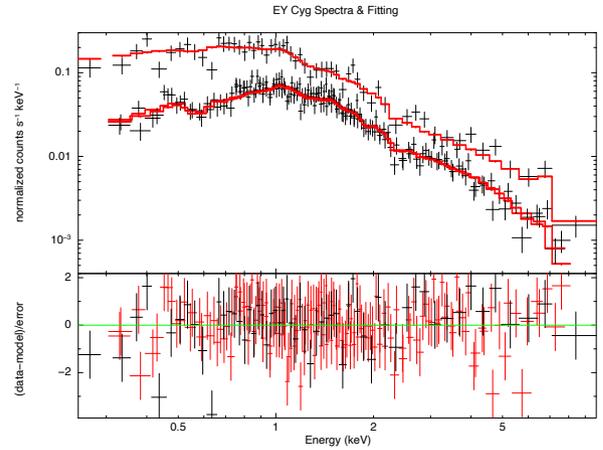}
\caption{\label{fig:eyspec} EY Cyg spectrum and best fit results
with the model in Table 5. The pn data are above the MOS ones, the
 model is in red, and the residuals have the same color key as in Fig.7. 
}
\end{figure}

\citet{Mukai2003} have argued that a cooling flow model fits well the X-ray
 spectra of disk accretor CVs, and we found that 
the XSPEC model vmcflow fits the spectra reasonably
 well,  but so does also a two-temperature plasma model (two APEC regions). 
The spectrum and the spectral fit with the cooling flow
is shown in Figure \ref{fig:eyspec} and the 
best fit parameters are in Table \ref{tab:eyspec}.
\begin{table}
\centering
\caption{\label{tab:eyspec}Best fit parameters for EY Cyg with a cooling flux model.}
\begin{tabular}{|c|c|} \hline
Parameter & Value \\ \hline
N(H) (TBABS, $\times10^{22}cm^{-2}$) & 0.028$_{-0.011}^{+0.012}$ \\ \hline
 low T (keV) & 0.09 (Fixed) \\ \hline
high T (keV) & 23.54$_{-4.8}^{+6.8}$ \\ \hline
$\dot{m}$ ($\times10^{-11}M_\odot/yr$) & 1.77$_{-0.28}^{+0.32}$ \\ \hline
$F_{\rm abs}$ ($\times10^{-12}erg/cm^2/s$) & 1.34$_{-0.16}^{+0.12}$ \\ \hline
$F_{\rm unabs}$ ($\times10^{-12}erg/cm^2/s$) & 1.44 \\ \hline
$\chi^2$ & 1.15 \\ \hline
\end{tabular}
\end{table}

In the cooling flow model, $\dot m$ is a fit parameter. We obtained
 the best fit with $\dot m= 1.77 \times 10^{-11}$ M$_\odot$ year$^{-1}$
and the value derived from the fit with two
 plasma temperatures is about the same.
Also the values of column density (N(H)) and the fluxes are about the same in the two different fits. 
\subsection{EY Cyg: timing analysis}
Because the background was enhanced, but it was not extremely elevated during much of the exposure of this target, we could correct the light curve for the background during the whole exposure and used all the exposure time of each of the three cameras.  As shown in Fig. 10, 
in the Lomb Scargle
 periodogram of EY Cyg, we find a frequency of 0.205$^{+0.12}_{-0.11}$ mHz
 with 99.9\% probability. It corresponds
 to a period of 4867$^{+275}_{-262}$ s, or 
 about 81 minutes, which is very different from the orbital period measured
 by 
Echevarria et al. (2007). If it corresponds to the WD rotation period, this leaves the possibility open that EY Cyg is an IP 
 like V2491 Cyg. We also show the light curve folded with this period in the second panel of Fig. 10.
In the Lomb-Scargle periodogram in Fig. 10 we also notice another
 the frequency at $\simeq$0.4 mHz, which is close to the first harmonic
 of the main signal. We also experimented by splitting
 the light curve into two halves:
 in the light curve of the second half of the exposure, we
 found a significant frequency at $\simeq$0.27 mHz, but it may have 
 been a transient event\footnote{Dividing the light curve into many short subsamples showed this peak only in a very short time interval. However, such short light curve subsamples yield very low resolution of the corresponding periodograms, therefore not suitable for any further detailed study. } or an artifact.

 We also found an orbital modulation,
 as also noted by \citet{Nabizadeh2020}, although the exposure was not sufficiently long
 to detect the corresponding frequency.
 The pn and MOS-2 light curve is shown folded over the orbital period in Figure \ref{fig:eylc}. We also show the light curve in selected
 MOS-2 energy bands: 0.3-0.6 keV, 0 and 1-10 keV. The
 1-10 keV band seems less modulated, and because we could only follow one orbital period, the high values may be attributed to a flux increase that is not orbital in nature. 
\citet{Nabizadeh2020} attributed the modulation mainly to the secondary star.
 A contribution of the disk 
appears unlikely at such low inclination, while it is
 reasonable that the modulation is indeed due to the contribution
 of the companion because K stars emit all of their
 X-ray flux below 1 keV: Fig. 11 shows that the modulation
 is smooth at the softest energy.  
On the other hand, the flux
 below 0.6 keV is only about 15\% of the total flux and does not
 ``pollute'' the accretion disk boundary layer's flux significantly,
 so our conclusion on $\dot m$ {remains valid}. 
\begin{figure*}
\centering
\includegraphics[width=0.49\linewidth]{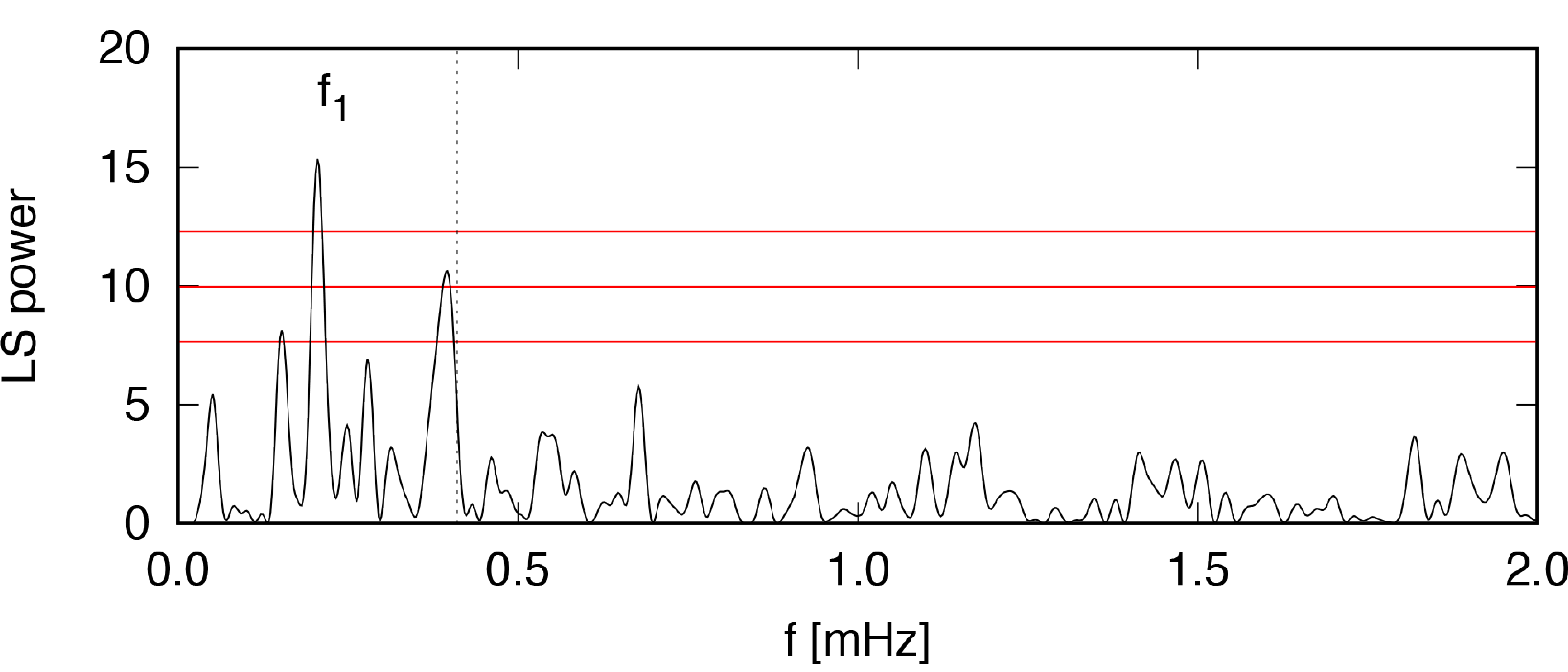}
\includegraphics[width=0.49\linewidth]{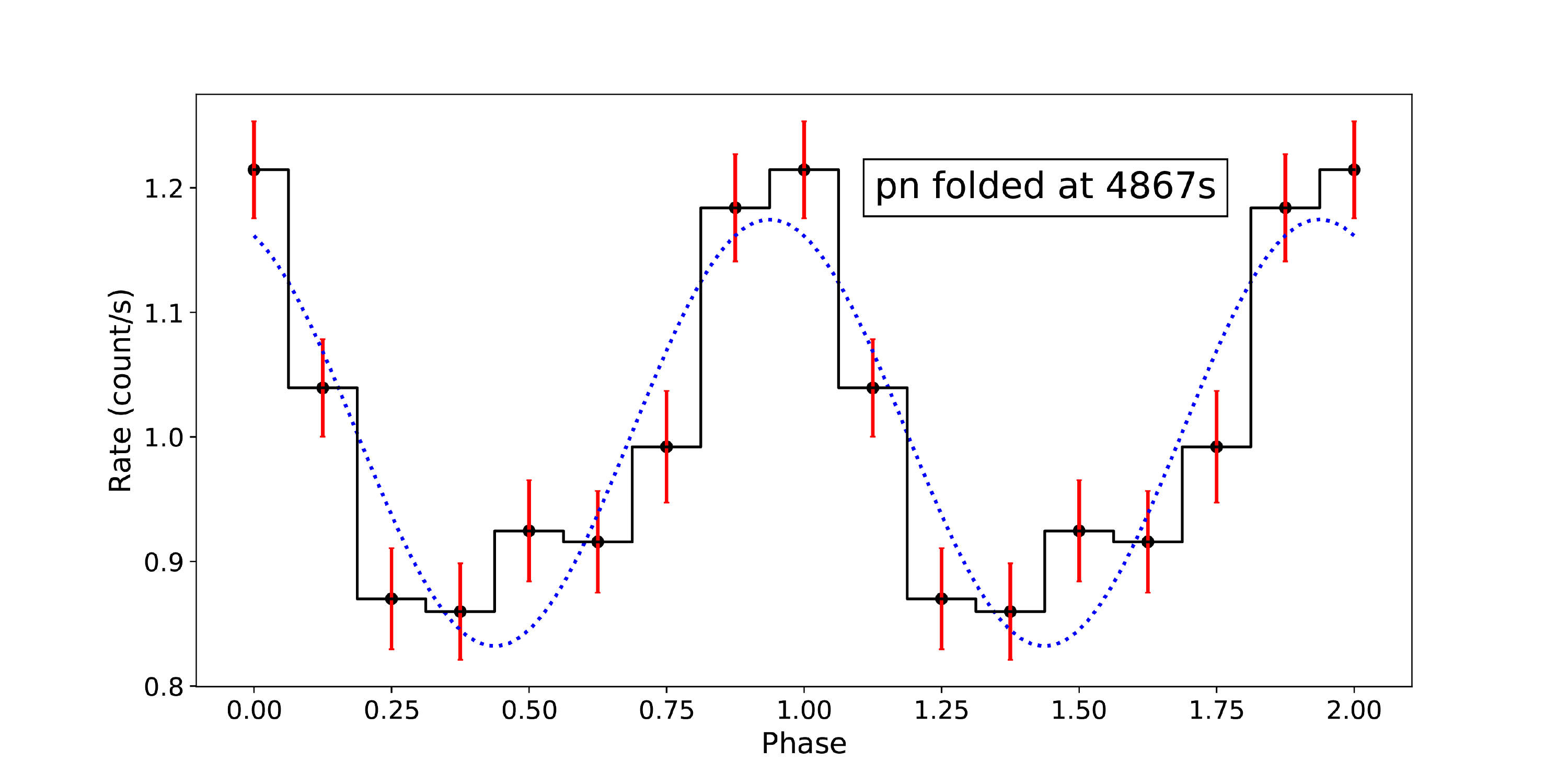}
\caption{{\bf The first panel shows} the Lomb-Scargle periodogram for the
 detrended pn light curve of
 EY Cyg in which the red lines represent the 99.9\%, 90.0\% and
 80.0\% confidence levels, and the vertical dashed lines shows the 
 first harmonic of $f_1$. 
 The second panel shows
the pn light curve in the 0.2-10 keV range folded with the 4860 s period.}
\end{figure*}
\begin{figure}
\centering
\includegraphics[width=0.5\textwidth]{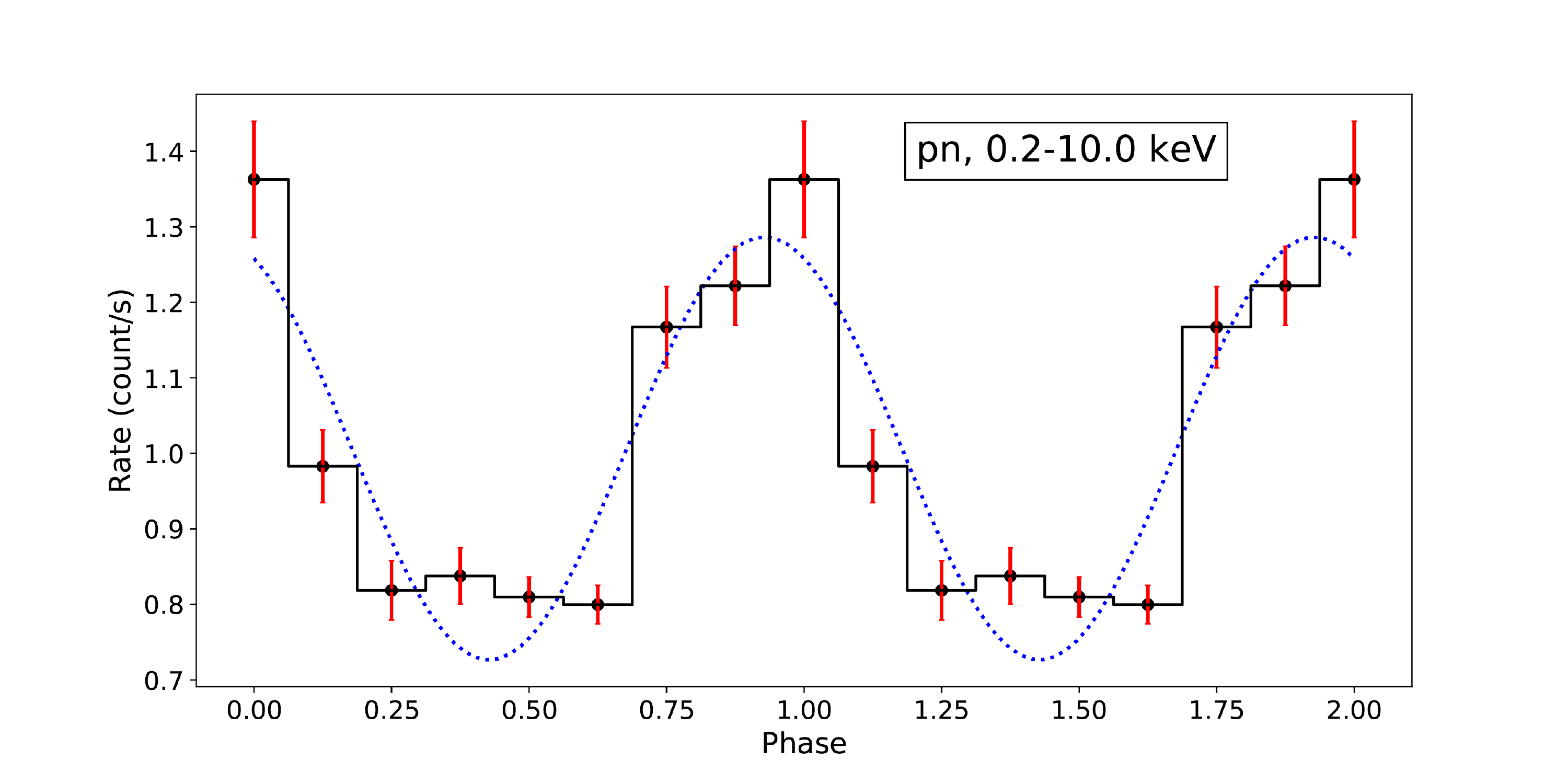}
\includegraphics[width=0.5\textwidth]{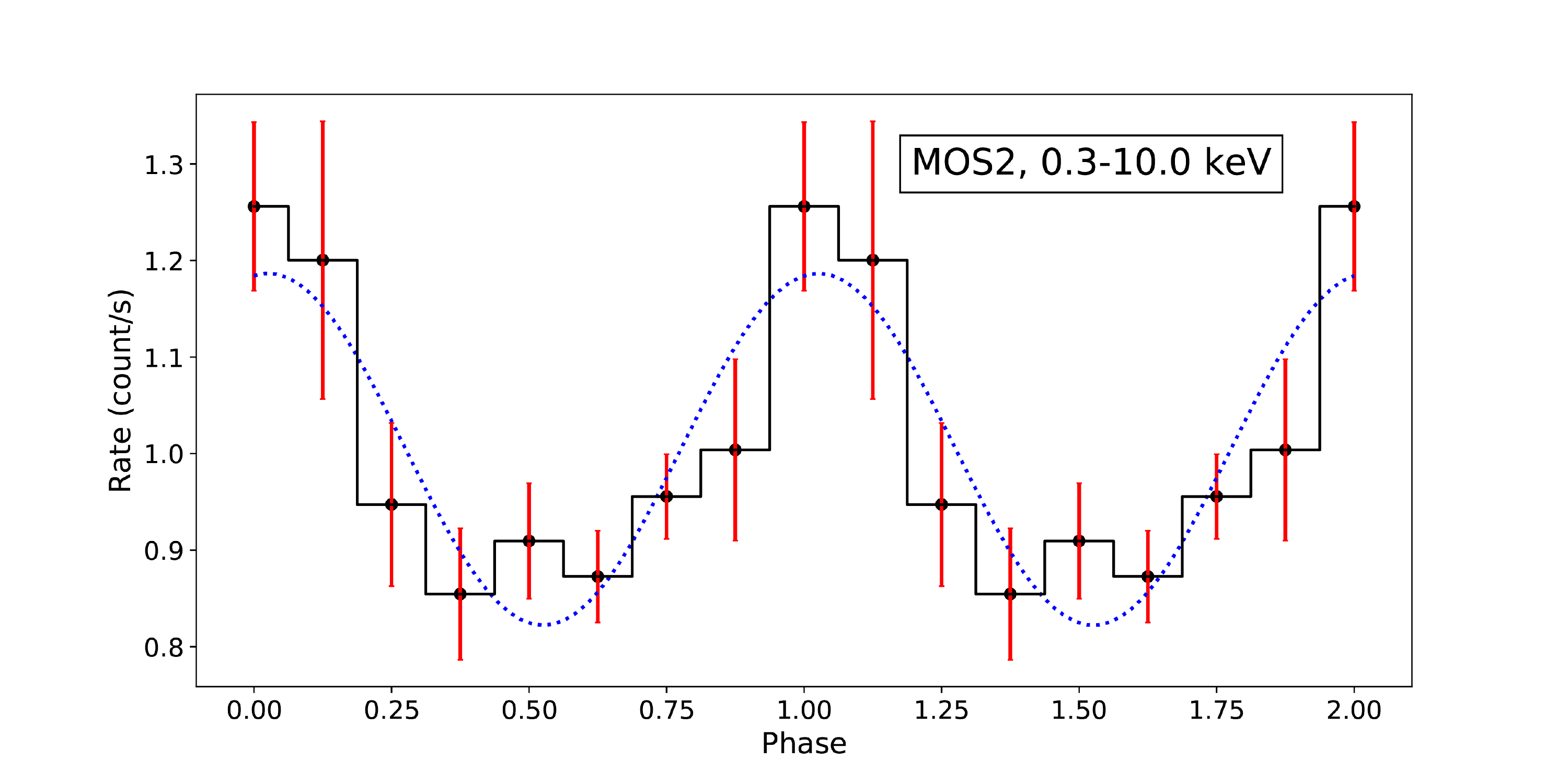}
\includegraphics[width=0.5\textwidth]{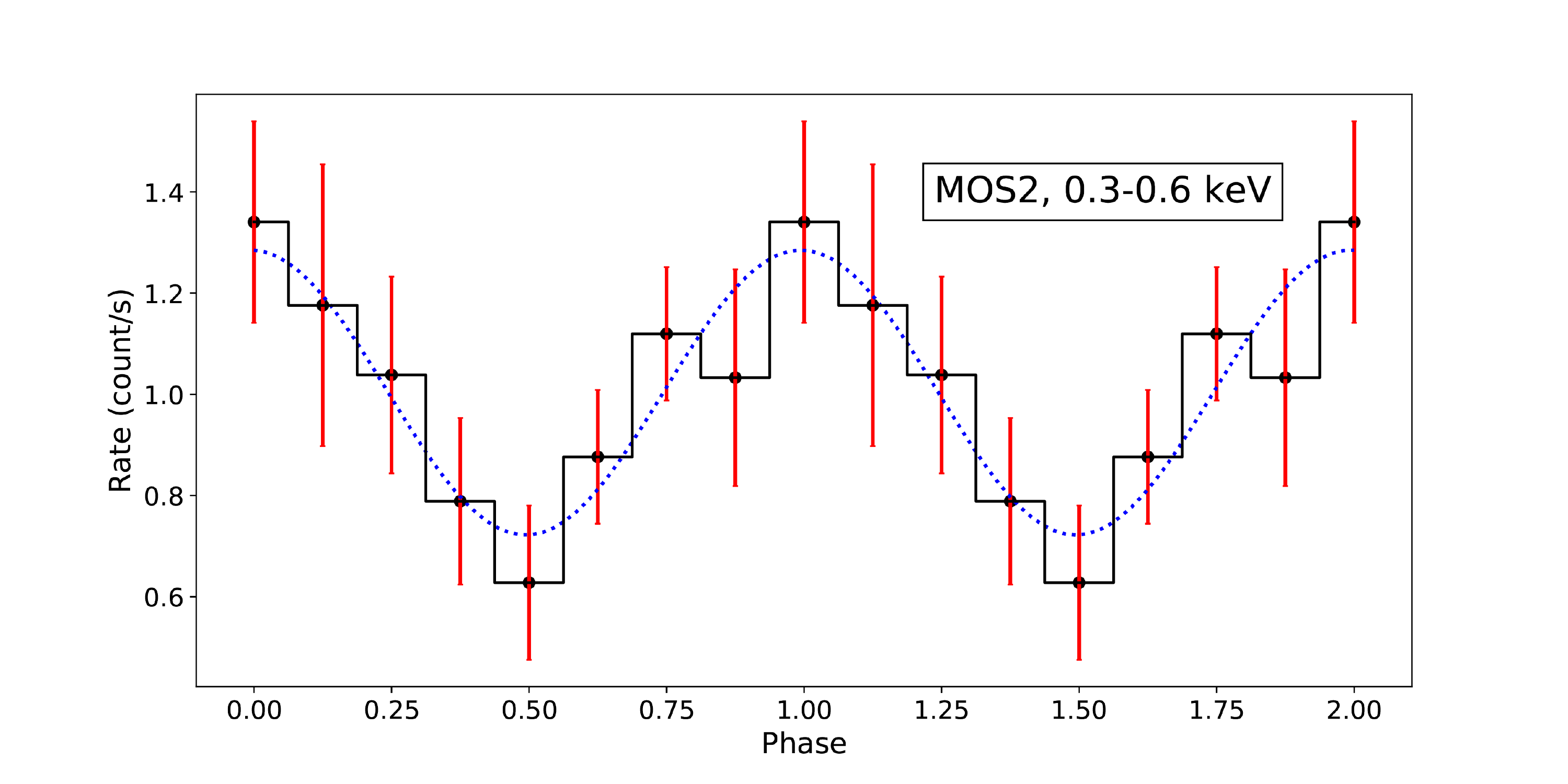}
\includegraphics[width=0.5\textwidth]{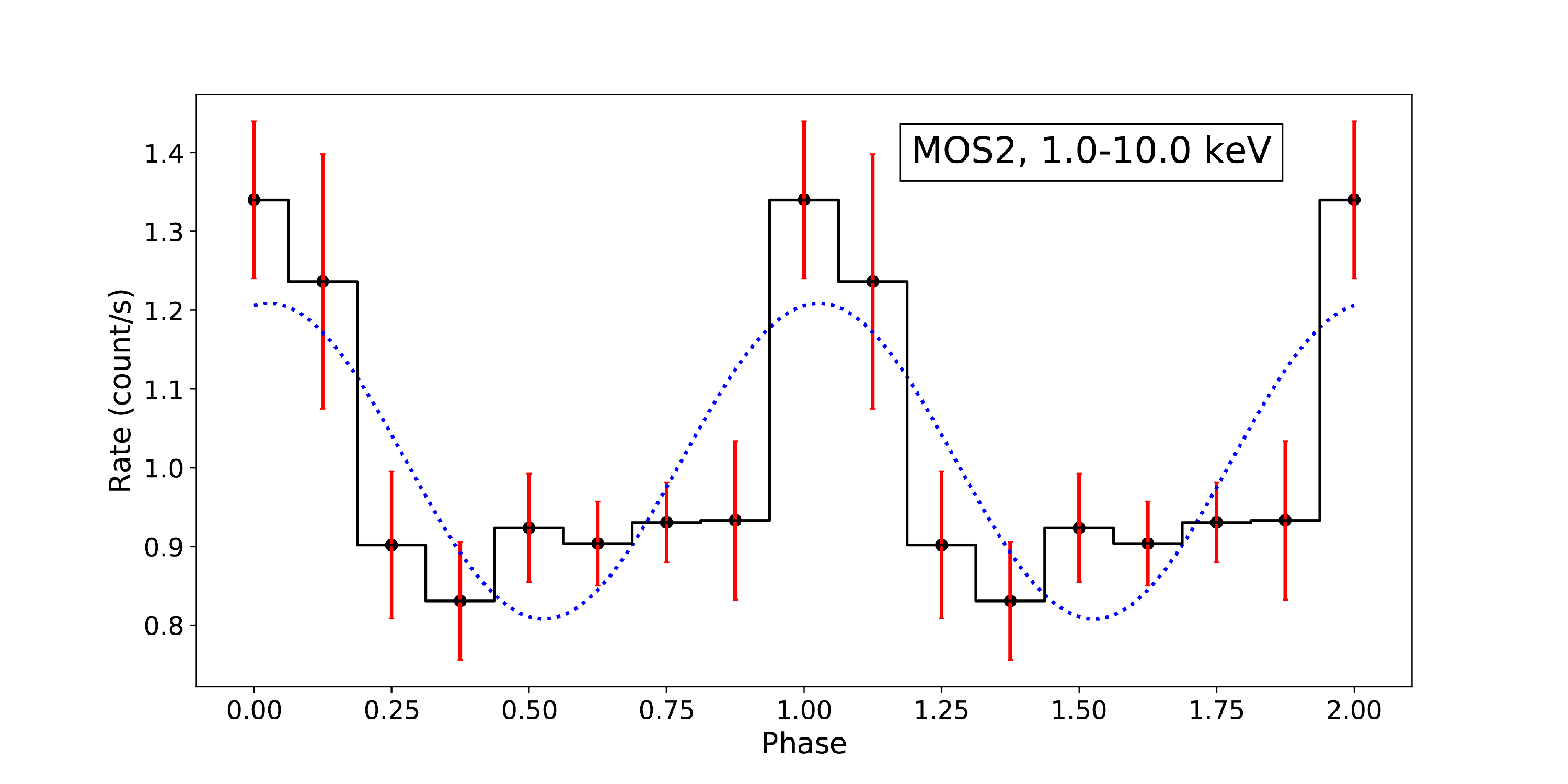}
\caption{\label{fig:eylc} Light curves of EY Cyg  folded with the orbital period,
for the pn and for the MOS2. For the MOS2 we also present the very soft (0.3-0.6 keV)
 and the hard (1-10 keV) ranges.  The count rate has been normalized
 to the average of each light curve, and
the blue dots show a sine wave fit to the folded light curve, which assists to visualize the periodicity. }
\end{figure}
\subsection{V794 Aql}
The medium energy Chandra High Energy Transmission Gratings (HETG) were used for the V794 Aql observation, specifically the Medium Energy Gratings (MEG) and the High Energy Gratings (HEG), with a respective absolute wavelength accuracy of 0.0006 and 0.011 Angstrom. 
 The observation was done on 2015 August 25 for 94.5 kiloseconds. {Because this target is not in the AAVSO or other public databases for variables, we obtained and scheduled optical observations for V794 Aql on a date as close as possible to the one of the Chandra exposure. }

We monitored
 the target optically on 2015 August 30 of the same year with the 60cm
 Cassegrain telescope in Stara Lesna, Slovakia, 
 and in R also with the 18cm Maktsutov site on the same site; 
 then again on 2015 September 10 with the 1m Cassegrain telescope 
 at the Special Astronomical Observatory of the Caucasus in Russia.
 Despite some small
 amplitude flickering and occasional small flares,
 the star's luminosity does not vary significantly
 in magnitude over times of few hours, and we
 report here the average magnitudes in the Johnson filters.
 On 2020 August 30 we measured R=14.866$\pm$0.015 in 11 exposures
 taken within 17 minutes, V=15.160$\pm$0.021 in
%V=15.136  B=15.324
  6 exposures taken within 11 minutes, B=15.229$\pm$0.050 in 2 
 B exposures within 3 minutes. On 2020 September 10 the average magnitudes
 were R=14.627$\pm$0.012 in 17 exposures taken
 within 2.86 hours, V=14.737$\pm$0.012 in 18 exposures taken
 during 2.8 hours, B=14.921 in 17 exposures taken within 2.68 hours.   
\citet{Honeycutt2014} showed that in the high states, V794 Aql varies between
 V=14 and V=15.5, with amplitude variations within a few days; occasionally
 (every few years) there is a low state at V$\simeq$17.5-18,
 that lasts for over a month and takes a mean time
 of 23$\pm$4 days for the decay to the minimum, and of 11$\pm$5 days for the rise
 back to maximum. Despite the variations in the 10 days between optical
 observations, this interval is too long, not typical of the sharp rise 
 observed after previous low states. The 5 days between the
 {\sl Chandra} observation and the first optical measurement,
 on the other hand, would not have been enough for a full rise
 back to maximum.  We also have to consider that   
 no low states have been reported since 2011, and before
 2007 there was a previous
 11 years interval without low states \citep{Honeycutt2014}.
 Thus, it is extremely likely that the {\sl Chandra}  exposure,
 which is less than 5 days before the optical
 one was done during the more common optically high state of V794 Aql. 

The {\sl Chandra} data were extracted with the CIAO
software \citep{Fruscione2006} version 4.9.1 and the CALDB calibration package version 4.8.3. 
The spectra were fitted with XSPEC. 
 Unfortunately, these spectra do not have the high S/N we had anticipated.
 V794 Aql was observed with about the same X-ray flux as in the low state measured by \citet{Zemko2014}. Even in a previous intermediate state, the 
X-ray flux was four times larger. 
  The spectrum appears to be very hard and to peak outside the
 {\it Chandra} ACIS+HETG range, so that we
 are mostly probing the low energy tail of the X-ray flux. 
 Only emission lines at energy of at least 
 1.5 keV (8.42 \AA) are measured.
 Fig. 11  shows our identification of a strong H-like line
 of Mg XII at 8.42 \AA, while Mg XI He-like lines are not observed.
 We identified
 lines of S XVI, Si XIV, and Si XIII that were also strong in TT Ari \citep{Zemko2014},
 and we also tentatively identify as the H-like line of Ca XX. We fitted
 the HEG and MEG spectra simultaneously because we had few
 counts per bin. Since we did not want to use a high binning factor such that would
 loose information on the emission lines, we used Cstat
 statistics \citep{Cash} to find the best fit that minimizes the Cstat
 parameter. Several simple models can all fit the data, albeit
 not perfectly: 
 an APEC plasma in collisional ionization equilibrium, 
 two APEC components at different temperature with solar abundance, and
 the MKCFLOW model described above.  
 We show the parameter of the latter in Table 7 and the fit in Fig. 11.
 The maximum temperature
 is not well constrained and reaches the maximum value of the model, 72 keV,
 but at the 90\% confidence level uncertainty, it may be as low as 43 keV.
 The predicted $\dot m$ is 1.5 $\times 10^{-11}$ M$_\odot$ yr$^{-1}$. 
 In the two APEC component models 
 the maximum temperature also reaches the maximum value (64 keV in this case) and
 maybe as low as 17 keV with the 90\% confidence level uncertainty.
 We cannot fit the spectrum with only one temperature; in fact
 the strong Mg XII line is produced at low temperature, and the continuum
 indicates a much higher temperature. 
\begin{figure}
\centering
\includegraphics[width=0.45\textwidth]{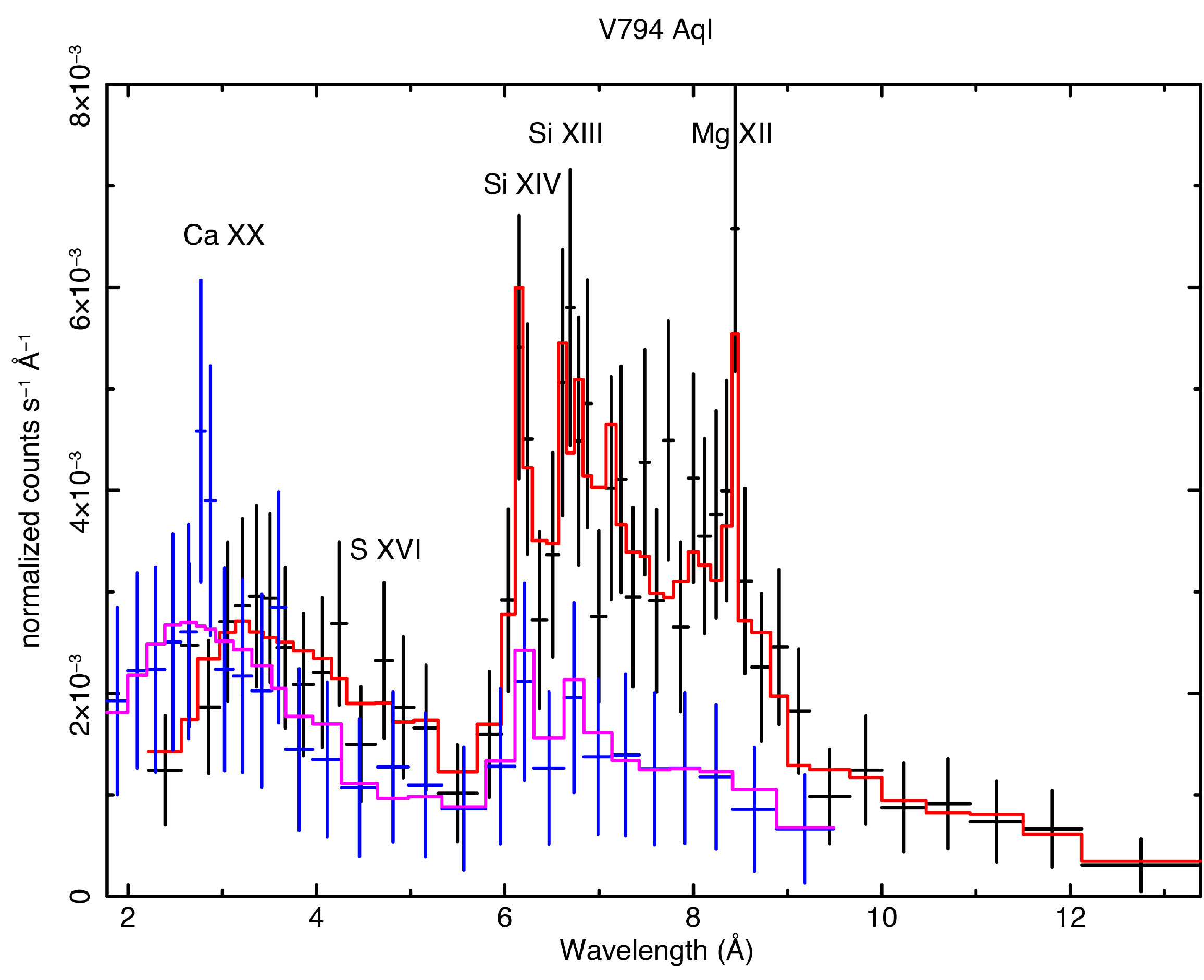}
\caption{\label{fig:v794spec} The HEG (in blue) and the MEG (in black) spectra
 of V794 Aql and the fit with Model 1. (in red for the MEG and pink for the HEG). 
}
\end{figure}
\begin{table}
\centering
\caption{Best fit parameters for the V794Aql HEG and MEG spectra, fitted simultaneously
 using the MKCFLOW model in XSPEC. The maximum and minimum
 temperature have reached the last value included in the XSPEC model.}
\begin{tabular}{|c|c|}
\hline 
Parameter & Value \\
\hline
N(H) ($\times10^{21}$ cm$^{-2}$) & 2.7$^{+1.3}_{-0.6}$ \\
\hline
T$_{\rm low}$ (keV) & 0.08$^{+0.10}$ \\
\hline
T$_{\rm high}$ (keV) & 72$_{-39}$  \\ \hline
Z  & 1.8$^{+1.2}_{-0.8}$  \\ \hline
$\dot m$ &    2.1$^{1.6}_{-0.3}$\\
\hline
F$_{\rm abs.}$   & 3.0$^{+2.0}_{-1.3} \times 10^{-12}$   \\ \hline
F$_{\rm unabs.}$ & 3.4 $\times 10^{-12}$  \\ \hline
\end{tabular}
\end{table}
\section{Discussion}
\subsection{Similar in outburst, different in quiescence}
 V2491 Cyg and KT Eri share similar characteristics in being fast novae with an outburst optical amplitude of about 9 mag, 
at the low end for CN and closer to RN's typical amplitude. 
In X-rays, however, {these two
 systems} are very different at quiescence. 
V2491 Cyg is most likely an IP accreting at a high rate;
 KT Eri instead does not have signatures of a magnetic system. 
Assuming the reasonable hypothesis that
 the X-ray luminosity of the two recent novae in our sample is due to accretion, V2491 Cyg is accreting at
 two orders of magnitude higher rate than KT Eri. This is consistent with
 V2491 Cyg having a short recurrence time, perhaps as short as an RN, while
 either KT Eri has entered a phase of ``hibernation'', or its outburst
 recurrence time is greater than a thousand years \citep[see model calculations of][]{Yaron2005}.
 The supersoft X-ray flux of V2491 Cyg was still very high at the end of 2017 and 
  is not clearly associated with accretion. It may be due to a hot polar cap under which nuclear burning is still continuing or has just been quenched. On the other
 hand, the very soft portion of the KT Eri
 flux has much lower X-ray luminosity and is consistent with
 accretion as underlying mechanism. Most likely, V2491 Cyg is an IP and
 KT Eri is a disk accretor. 
 
\subsection{V2491 Cyg as an IP: the WD rotation period}
One of our most important findings is that a clear 39 minutes periodicity
 of V2491 Cyg is still measured and corresponds to the previously
 measured frequencies within the statistical uncertainty of the data.
Thus it is very likely to be due to the WD rotation. 
 The periodicity is not direct proof of
 the WD magnetic field, but it is strong evidence in favor of the
 classification of this nova as an IP.
We also measured another frequency, about half of the known one, namely $f_1$.  This may imply that $f_1$ is
 the true rotational frequency, while the clear detection
 of $f_2$, corresponding to the $\simeq$39 minutes period, maybe 
an effect of two polar regions at different temperature and X-ray
 luminosity. The upper panel of Fig.~\ref{lc_binned} shows
a double-humped structure, with two humps at different amplitudes.
 If the amplitude was almost the same, 
the period analysis method would ``see'' only the shorter
 period modulation $f_2$. One hump would be detected as one cycle, while 
we would not be able to identify the longer ($f_1$) cycle.
 This may have been the case of the two previous {\it XMM-Newton} and {\it Suzaku}
 light curves, where the peak at $f_1$ did not appear. In
 our new data, the two humps have different amplitude, and
 the analysis ``distinguishes'' the two humps, both 
 periodically repeated at corresponding frequency $f_1$, 
so  this peak appears in the periodogram, but
 the folded light curve with the higher $f_2$ frequency does not shows 
this structure (lower panel of Fig.~\ref{lc_binned}).
A similar double-humped structure and two frequencies were
 detected in the X-ray observations of the IP
V2069\,Cyg \citep{Butters2011}, where the modulation is rotational
 in nature. The physical reason for V2491 Cyg may be the different
luminosities of the accretion regions at the two poles,
resulting in different maxima of the humps in the folded light curve.  
If the difference in luminosity between the two poles was not so large
 yet during the outburst and at the beginning of quiescence, this would
 explain why we detected only the $f_2$ frequency. 

 In two other IP novae, V4743 Sgr
 and V407 Lup, the supersoft X-ray flux of the luminous central
 source before the end of nuclear burning decreased at almost constant
 temperature \citep{Zemko2016, Aydi2018}. This has been interpreted as
 an indication of a hot and luminous shrinking region on the WD,
 most likely consisting of the polar caps on which fresh matter is accreted,
 either keeping the high T$_{\rm eff}$ for longer, or fueling
 continued nuclear burning {that remains confined
 to the deeper layers and does not expand} 
 in the rest of the surface. 
 In V2491 Cyg we still detected large supersoft flux: possibly, 
 this effect of polar accretion was still continuing.

\subsection{X-ray luminosity probing accretion rate}  
Despite several early findings that the X-ray  flux of CVs may be too low to
be due to energy dissipated in the accretion process \citep[e.g.][and
 references therein]{Mauche1998,
Orio2001}, 
 the X-ray luminosity of V2491 Cyg is consistent with high
 accretion rate predicted for a nova of low amplitude and with similarities to
 RNe \citep[see calculations by][]{Yaron2005}.
KT Eri is at much lower X-ray luminosity, and the derived $\dot m$ is
 not consistent with the models' predictions for a nova with an optical amplitude of only 9 mag \citep{Yaron2005}.  
 All the outburst models, in fact, generally {indicate} that the amplitude
 is larger if the nova accretes at low $\dot m$. This can be reconciled
 with what we observe only of this relatively small amplitude
 is caused by an optically luminous, evolved secondary.

We find that the X-ray luminosity of the dwarf nova and ex-CN candidate,
 EY Cyg, is consistent with the low $\dot m$ expected for dwarf novae
 and also, most likely, for ``hibernating'' novae. 
Recently \citet{Nabizadeh2020}, shortly after
 our presentation at the 2020 AAS winter meeting \citep{Sun2020}, also examined our 
 EY Cyg data, available in the HEASARC archive. These authors
favor a multi-temperature plasma emission model (CEVMKL),
 analogous to VMFLOW if the power of the emission measure distribution 
 is consistent with radiative cooling.
  Their fit indicates hydrogen column density
 $N_H=0.04_{-0.01}^{+0.02}\times10^{22}cm^{-2}$, which is consistent with our
 result in Table \ref{tab:eyspec}, but 
their result on the unabsorbed X-ray flux is $19\pm 1 \times 10^{-13}$
erg cm$^{-2}$ s$^{1}$ in the 0.1-50.0 keV range. This flux is larger than what
 we obtain extrapolating the flux to 50 keV and would indicate an
 even larger $\dot m$ than we obtained using Equation 1.
 However, \citet{Nabizadeh2020} derived  $\dot{m}=6.0^{+0.3}_{-0.3}\times10^{-12}$ M$_\odot$ yr$^{-1}$,
which is only approximately half of what we estimated with
 the assumption that half of the gravitational energy is emitted
 as X-ray light.
 We disagree with these authors' conclusion that EY Cyg is
 underluminous in comparison with what is expected from the boundary layer of the disk and that invoking an advective flow is necessary to explain
 the data; we rather find that the $\dot m$ value is 
 consistent with an approximate estimate obtained form the UV flux.
There was no rigorous
 determination from the UV flux in \citet{Sion2004}, 
{in fact} the given value can be considered only as upper
 limit.

Although we agree with \citet{Nabizadeh2020} 
in attributing the soft portion of the X-ray flux comes from the
 K star, the soft flux below 1 keV is only 20 times lower than that of the harder portion, thus only borderline relevant in
contaminating the result of the emission from accretion
if it represents the X-ray flux of the secondary star.
 In the spectral range  previously
 evaluated, $\sim$K5-M0 (late K to early M)
 \citep{Connon1997}, the X-ray luminosity is 10$^{29}$-10$^{30}$ erg s$^{-1}$,
 but for K0 stars \citep{Nabizadeh2020} quote a case in which
 the luminosity reached even $\simeq 10^{31}$ erg s$^{-1}$. 
Thus the first APEC component in our Model 2 in the table, which has a temperature of 0.93 keV,
 may represent the (usually quite soft) contribution of the K star,
but it yields only an absorbed flux of $5.5\times10^{-14}$
erg cm$^{-2}$ s$^{-1}$ and an unabsorbed flux of $6.5\times10^{-14}$
 erg cm$^{-2}$ s$^{-1}$. {This is negligible compared to the total X-ray flux, the bulk of which seems to be due to accretion. }

\subsection{Are the VY Scl binaries nova-like accreting at high $\dot{m}$ during the whole high states period?}
One surprising result we obtained is the apparent lack of any clear
 correlation between the high and the low optical states of 
 the nova-like VY Scl binary
 of our sample, V794 Aql,
 and its X-ray flux level. Moreover, the X-ray spectrum of V794 Aql
 is very hard: 
 emission lines due to atomic transitions at lower energy
 than 1.5 keV are absent. If the X-ray flux is indicative of accretion, a VY Scl
 binary like V794 Aql does not only have decreased $\dot m$ in the optically
 low states, but accretion perhaps is intermittent. This is a likely
 reason for which nuclear burning, with the manifestation of a luminous supersoft X-ray source, is not ignited \citep{Zemko2014}.

 The plasma temperature of the hotter component in this nova-like exceeds 33
 keV. It is one of the ``hardest'' CVs, and such a high temperature almost certainly indicates a massive WD, although we could not constrain the plasma temperature well. A more definite result
 may be obtained with the {\sl NuSTAR} satellite, with the
 3-79 keV bandpass of its X-ray telescope. 
\section{Conclusions}
An important motivation to obtain a second exposure of both
 V2491 Cyg and KT Eri as they ``settled'' into quiescence was to
 monitor the supersoft X-rays, coming from  a region that was apparently
 shrinking at a constant temperature. In V2491 Cyg, we found that this region is
 still luminous and hot. In KT Eri, we witness the hardening of the 
X-ray spectrum over a few years, but
 our analysis of the previous data obtained with
 {\sl Chandra} indicates that the luminosity was already so low that it may
 have nothing to do with lasting heat on the polar caps due to
 long lasting nuclear burning underneath them. 
 KT Eri does not show any signs of magnetic
 accretion, and is probably accreting only through a disk. 

 We derived high accretion rate, above 10$^{-9}$ M$_\odot$ yr$^{-1}$
 for V2491 Cyg, and much lower, about 2 $\times 10^{-10}$ M$_\odot$ yr$^{-1}$ 
 for KT Eri.
 We confirmed that V2491 Cyg has the characteristics
 of an IP, and we found an intriguing $\simeq$81 
minutes period in the X-ray light curve of EY Cyg, which leaves open 
 the possibility that is another IP, although the high background of the
 exposure makes the result still uncertain.

  An orbital modulation of the X-ray flux in EY Cyg, a low inclination 
 system, may be due to the softest X-ray portion coming from the
 secondary. The $\dot m$ value, although low enough to be consistent
 with its dwarf nova outbursts, is not inconsistent with the UV luminosity
 attributed to the accretion disk.

  We found that V794 Aql is a hard X-ray emitter, indicating a high
 mass WD, but we also discovered that $\dot m$ was quite lower
 than previously measured during optically high states. We concluded that $\dot m$ is probably very variable even during the high states. In the light of the recent nova outburst of a VY Scl nova-like system, these large fluctuations in $\dot m$ may also be a significant feature of nova systems, rather than a cyclic process of short periods of high $\dot m$ with ``hibernation'' in between.

\section*{Data Availability}
The data analyzed in this article are all available in the HEASARC archive of NASA at the following URL: \url{https://heasarc.gsfc.nasa.gov/db-perl/W3Browse/w3browse.pl}

\section*{Acknowledgements}
M.O. is grateful to Koji Mukai for very useful discussions and help in the 
 interpretation of the V794 Aql spectrum.
M.O. also acknowledges NASA funding for the analysis of the XMM-Newton observations of EY Cyg, V2491 Cyg and KT Eri, and
 a  Chandra award for V794 Aql. 
 Support of an Italian  INAF award dedicated to the search
 and study of targets that may be observed
 in the future with the Square Kilometer Array (SKA) and the Cerenkov
 Telescope Array (CTA) (``PRIN-INAF 2017:Towards the SKA and CTA era: discovery, localization, and physics of transient objects'')
 has helped putting this research together and making comparisons. 
A.D. was supported by the Slovak grant VEGA 1/0408/20. 
G.J,M.L. is a member of the CIC-CONICET (Argentina) and 
acknowledges support from grant ANPCYT-PICT 0901/2017.
S.S.'s participation
 was supported by the Program of Development of M.V. Lomonosov Moscow State
 University ``Leading Scientific Schools'',
 project ``Physics of Stars, Relativistic Objects and Galaxies'',
 by the Slovak Academy of Sciences grant VEGA No. 2/0008/17, and by 
the Slovak Research and Development Agency under the contract No. APVV-15-0458,
%%%%%%%%%%%%%%%%%%%%%%%%%%%%%%%%%%%%%%%%%%%%%%%%%%



\bibliographystyle{mnras}
\bibliography{sample} % if your bibtex file is called example.bib

%%%%%%%%%%%%%%%%%%%%%%%%%%%%%%%%%%%%%%%%%%%%%%%%%%

%%%%%%%%%%%%%%%%% APPENDICES %%%%%%%%%%%%%%%%%%%%%

% Don't change these lines
\bsp	% typesetting comment
\label{lastpage}
\end{document}